\documentclass[cp]{copernicus}
\newcommand\dg{\ensuremath{{}^\circ}}
\begin{document}\sloppy

\title{How unusual was autumn 2006 in Europe?}
\author{G.~J.~van Oldenborgh}
\affil{KNMI, De Bilt, The Netherlands}
\correspondence{G.~J.~van Oldenborgh (oldenborgh@\discretionary{}{}{}knmi.nl)}
\runningtitle{How unusual was autumn 2006 in Europe?}
\runningauthor{G.~J.~van~Oldenborgh}
\received{21 May 2007}
\pubdiscuss{8 June 2007}
\accepted{5 November 2007}
\published{21 November 2007}

\firstpage{1}

\maketitle
\begin{abstract}
The temperatures in large parts of Europe have been record high during the meteorological autumn of 2006.  Compared to 1961-1990, the 2m temperature was more than three degrees Celsius above normal from the North side of the Alps to southern Norway.  This made it by far the warmest autumn on record in the United Kingdom, Belgium, the Netherlands, Denmark, Germany and Switzerland, with the records in Central England going back to 1659, in the Netherlands to 1706 and in Denmark to 1768.  The deviations were so large that under the obviously false assumption that the climate does not change, the observed temperatures for 2006 would occur with a probability of less than once every 10000 years in a large part of Europe, given the distribution defined by the temperatures in the autumn 1901-2005.

A better description of the temperature distribution is to assume that the mean changes proportional to the global mean temperature, but the shape of the distribution remains the same.  This includes to first order the effects of global warming.  Even under this assumption the autumn temperatures were very unusual, with estimates of the return time of 200 to 2000 years in this region.  The lower bound of the 95\% confidence interval is more than 100 to 300 years.

Apart from global warming, linear effects of a southerly circulation are found to give the largest contributions, explaining about half of the anomalies.  SST anomalies in the North Sea were also important along the coast.

Climate models that simulate the current atmospheric circulation well underestimate the observed mean rise in autumn temperatures.  They do not simulate a change in the shape of the distribution that would increase the probability of warm events under global warming.  This implies that the warm autumn 2006 either was a very rare coincidence, or the local temperature rise is much stronger than modelled, or non-linear physics that is missing from these models increases the probability of warm extremes.
\end{abstract}

\begin{figure}
\vspace*{2mm}
\begin{center}
\makebox[0pt][l]{\large a}%
\raisebox{-1mm}{\includegraphics[width=\columnwidth,clip]{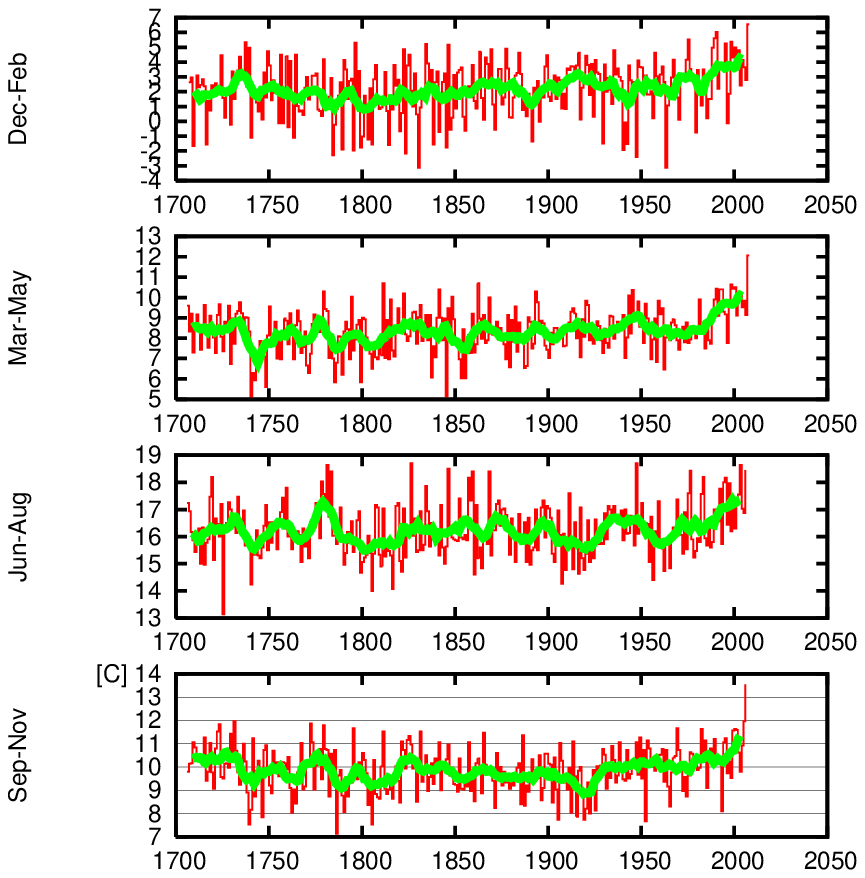}}
\makebox[0pt][l]{\large b}%
\raisebox{-1mm}{\includegraphics[width=\columnwidth,clip]{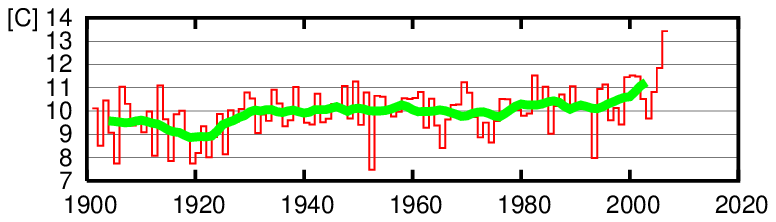}}
\end{center}
\caption{Autumn 2-meter temperatures at De Bilt, the Netherlands  with a 10-yr running mean (green) 1706-2006 \citep[a,][]{vanEngelenNellestijn1995}, 1901--2006 corrected for changes in the thermometer hut, location and city effects \citep[b,][]{Klimaatrapportage2003}.}
\label{fig:Tdebilt}
\end{figure}

\introduction
Meteorologically, the autumn of 2006 was an extraordinary season in Europe.  In the Netherlands, the 2-meter temperature at De Bilt averaged over September--November was 1.6\dg C \emph{higher than the previous record} since regular observations began in the Netherlands in 1706, which was a tie between 1731 and 2005 (Fig.~\ref{fig:Tdebilt}a).  The excess is much larger than the uncertainties in the earlier part of the record: \citet{vanEngelenGeurts1985} estimate that the standard error of monthly temperatures is 0.2--0.3\dg C \citep[see also][]{vandenDool1978}.  The Central England Temperature, 12.6\dg C, also was the highest since the beginning of the measurements in 1659, 0.8\dg C higher than the previous record of 1730 and 1731.  Pre-instrumental reconstructions indicate that September-November 2006 very likely was the warmest autumn since 1500 in a large part of Europe \citep{Luterbacher2007}.

The impacts of the high temperatures on society and nature have not been very strong, as in autumn a higher temperature corresponds to a phase lag of the seasonal cycle.  Flowering bulb farmers in the Netherlands were reported to have problems due to premature flowering.  However, a similar anomaly in summer would have given rise to a heat wave analogous to the summer of 2003, which caused severe problems \citep[e.g.,][]{SchaerJendritsky2004}.

In this article the heat anomaly of the autumn of 2006 in Europe is analysed.  First the observations are shown and return times are computed under the obviously false assumption of a stationary climate.  Next the first order effects of global warming are subtracted, and return times of the remaining weather signal computed.  The main weather factors are identified, and possible changes in their distribution are investigated using climate model simulations.

\section{Observations}

In Fig.~\ref{fig:Tdebilt} two time series of autumn (September--November) averaged temperature in De Bilt, the Netherlands are shown.  The first one covers 1706-2006 and is a combination of observations at various locations in the Netherlands converted to De Bilt temperatures \citep{vanEngelenNellestijn1995}.  The second time series 1901-2006 has been corrected for the effects of changes in thermometer hut, direct environment of the measurement location \citep{BrandsmavanderMeulen2007a,BrandsmavanderMeulen2007b}, and the effects of the growth of the cities in the area \citep{Brandsma2003}.  The value for 2006 is seen to be well outside the distribution defined by the other years in both series.

\begin{figure}
\vspace*{2mm}
\begin{center}
\includegraphics[width=\columnwidth]{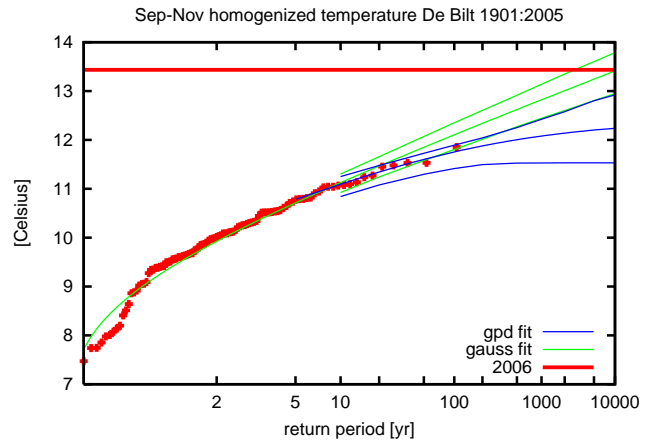}
\end{center}
\caption{Extrapolation of homogenised De Bilt autumn temperatures 1901--2005 (crosses) to the value observed in 2006 (horizontal line) using a fit to a Gaussian distribution and a Generalised Pareto Distribution (to the highest 20\%).  The return times have been computed under the obviously false assumption of only interannual variability.  The upper and lower lines indicate the 95\% CI, determined with a non-parametric bootstrap.}
\label{fig:Tdebilt_extra}
\end{figure}

Figure~\ref{fig:Tdebilt_extra} shows two extrapolations of the cumulative  distribution of the second, homogenised, time series 1901--2005, the first assuming a Gaussian (normal) distribution and the second using a peak-over-threshold method with a fit of the warmest 20\% to a Generalised Pareto Distribution \citep[GPD;][]{Coles2001}.  These extrapolations are based on the obviously false assumption that the autocorrelation is zero, i.e., that there are no long-term variations except those resulting from accumulations of interannual variability.  From figure~\ref{fig:Tdebilt_extra} it seems likely that the Gaussian distribution overestimates the probability of high temperatures.  In spite of this a return time of 10000 year is obtained with this extrapolation.  This shows that the assumption of no autocorrelation is very likely false: climate does change on longer time scales.  Physically, global warming has made high temperature anomalies much more likely during recent years, and this and other long-term variations increase the probability of clustered high extremes.

\begin{figure}
\vspace*{2mm}
\begin{center}
\includegraphics[angle=-90,width=\columnwidth,clip]{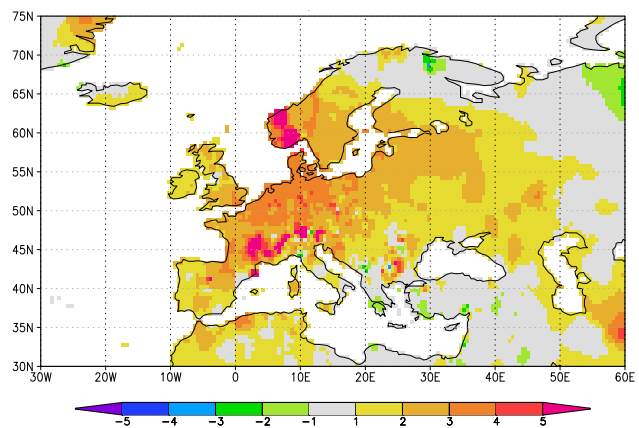}%
\end{center}
\caption{The temperature anomaly (degrees Celsius) relative to 1961-1990) of September--November 2006 in the GHCN/CAMS dataset \citep{GHCNCAMS}.}
\label{fig:ECAanomaly}
\end{figure}

\begin{figure}
\vspace*{2mm}
\begin{center}
\includegraphics[angle=-90,width=\columnwidth,clip]{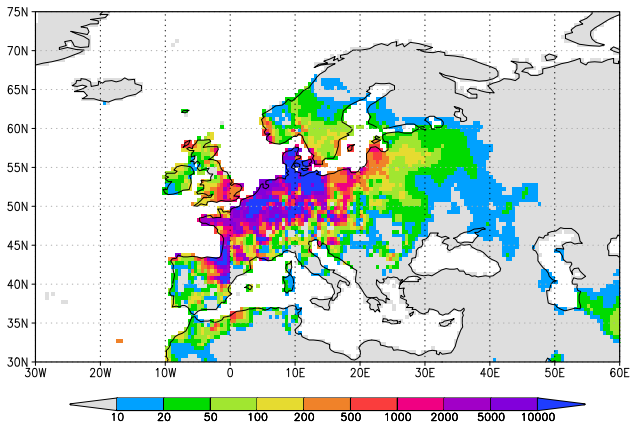}%
\end{center}
\caption{The return time (years) of September--November 2006 in the GHCN/CAMS dataset 1948-2005 under the false assumption of a stationary climate, using the more conservative Gaussian extrapolation.}
\label{fig:ECAgauss}
\end{figure}

The same analysis has been applied to all grid points of the 0.5\dg\ Global Historical Climate Network (GHCN) / Climate Anomaly Monitoring System (CAMS) \citep{GHCNCAMS}.  The GHCN and CAMS time series used in the construction of this dataset have not been homogenised, in contrast to the De Bilt series of Fig.~\ref{fig:Tdebilt}b (which is not included in this dataset).  All results have been verified against the 5\dg\ CRUTEM3 dataset \citep{CRUTEM3} (not shown).

Figure~\ref{fig:ECAanomaly} shows that the area with 3\dg C anomalies stretches from the the Alps to Denmark and from Belgium to Poland, with maxima in the the Alps and in southern Norway.  An extrapolation using a Gaussian fit (Fig.~\ref{fig:ECAgauss}) shows that in an unchanging climate the return times of this autumn would be 10000 years or more in an area shifted somewhat to the west of the area with highest anomalies.  (The shift is due to the smaller variability near the Atlantic Ocean.)  As can also be shown directly,  year-to-year autocorrelations are not negligible.

\section{Global warming}

\begin{figure}
\vspace*{2mm}
\begin{center}
\includegraphics[angle=-90,width=\columnwidth,clip]{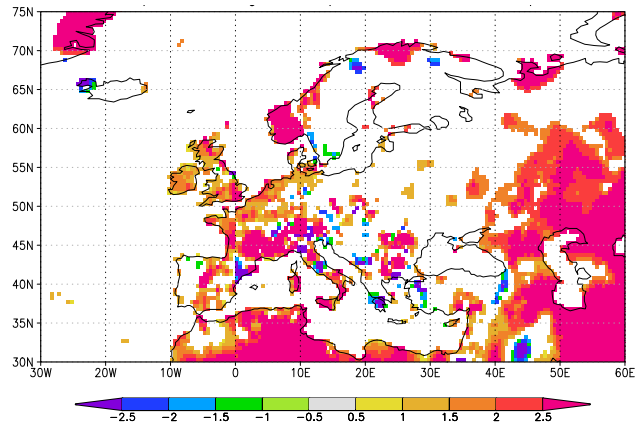}%
\end{center}
\caption{The regression $A$ of local against globally averaged temperature over 1948-2006.  Only grid points where the correlation is 90\% significant (assuming a normal distribution) are shown.}
\label{fig:A}
\end{figure}

The climate is not stationary: temperatures have been rising over Europe as in most of the rest of the world \citep{IPCC2007WG1}.  The effect of this on the probability of the occurrence of an anomaly like in autumn 2006 can be studied in a first approximation by subtracting the local temperature change proportional to a global temperature, here the 3-yr running mean of the global mean temperature, $T^{\prime(3)}_\mathrm{global}(t)$.  (On interannual time scales $T^{\prime}_\mathrm{global}(t)$ contains clear ENSO signals, and these should not enter in the description of how global temperature affects Europe, hence the low-pass filtering.)  This gives as a first approximation:
\begin{equation}
T'(t) = A T^{\prime(3)}_\mathrm{global}(t) + \epsilon(t)\,,
\end{equation}
where $\epsilon(t)$ denotes the part of the temperature not directly proportional to global temperature changes.  The coefficients $A$ are determined by a fit of local to global temperature up to 2005 and are shown in Fig.~\ref{fig:A}.

Station inhomogeneities (or local temperature anomalies)  are visible in the GHCN/CAMS dataset (these are not visible in the much coarser CRUTEM3 dataset).  On average the temperature in Europe has increased somewhat faster than the globally averaged temperature in autumn, in accordance with the Cold Ocean / Warm Land pattern \citep{Sutton2007}.  For the homogenised De Bilt time series, $A = 1.68\pm0.37$ ($1\sigma$ standard error) over 1901-2005.

Subtracting the local trend $AT^{\prime(3)}_\mathrm{global}(t)$ from the observed temperatures, the weather anomalies $\epsilon(t)$ remain.  The anomalies up to 2005 are described reasonably well by a Gaussian in autumn; the skewness is $-0.46\pm0.36$, so the Gaussian fit may even overestimate the positive tail.  The autocorrelation is now slightly negative at lags up to 4 year, so that consecutive years of this series can be considered independent.

\begin{figure}
\vspace*{2mm}
\begin{center}
\includegraphics[width=\columnwidth]{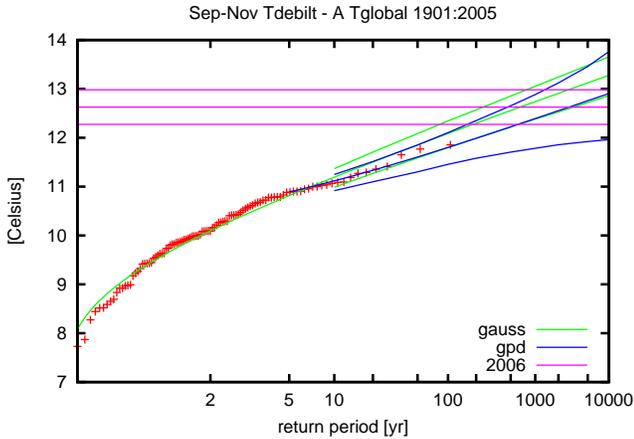}%
\end{center}
\caption{As Fig.~\ref{fig:Tdebilt_extra} but now for $T(t) - AT^{\prime(3)}_\mathrm{global}(t)$, the De Bilt temperature with the regression against the global mean temperature anomalies (with a 3-yr running mean) subtracted.  The 95\% uncertainty interval in the 2006 anomaly above the trend due to the uncertainty in the regression is denoted by the horizontal lines.}
\label{fig:Tdebilt-Tglobal_extra}
\end{figure}

In Fig.~\ref{fig:Tdebilt-Tglobal_extra} the cumulative distribution of the autumn temperatures minus the linear effect of global mean temperature changes 1901-2005 is shown.  Extrapolating this to the value observed in 2006 gives a return time of 650 years.  Using a GPD to fit the highest 20\% of the distribution gives a return time of about 3000 years.
To compute the statistical uncertainty margins in these numbers a non-parametric bootstrap was performed simultaneously on the regression and the extrapolation.  This gives a 95\% CI on the return time of 125 to 10000 years for the Gaussian extrapolation, 200 years to infinity for the GPD extrapolation.  The lower bounds are reached when the real trend is higher than the central fitted value of $A=1.68$, lowering the value for 2006 (lower horizontal line).

\begin{figure*}
\vspace*{2mm}
\begin{center}
\includegraphics[angle=-90,width=\columnwidth,clip]{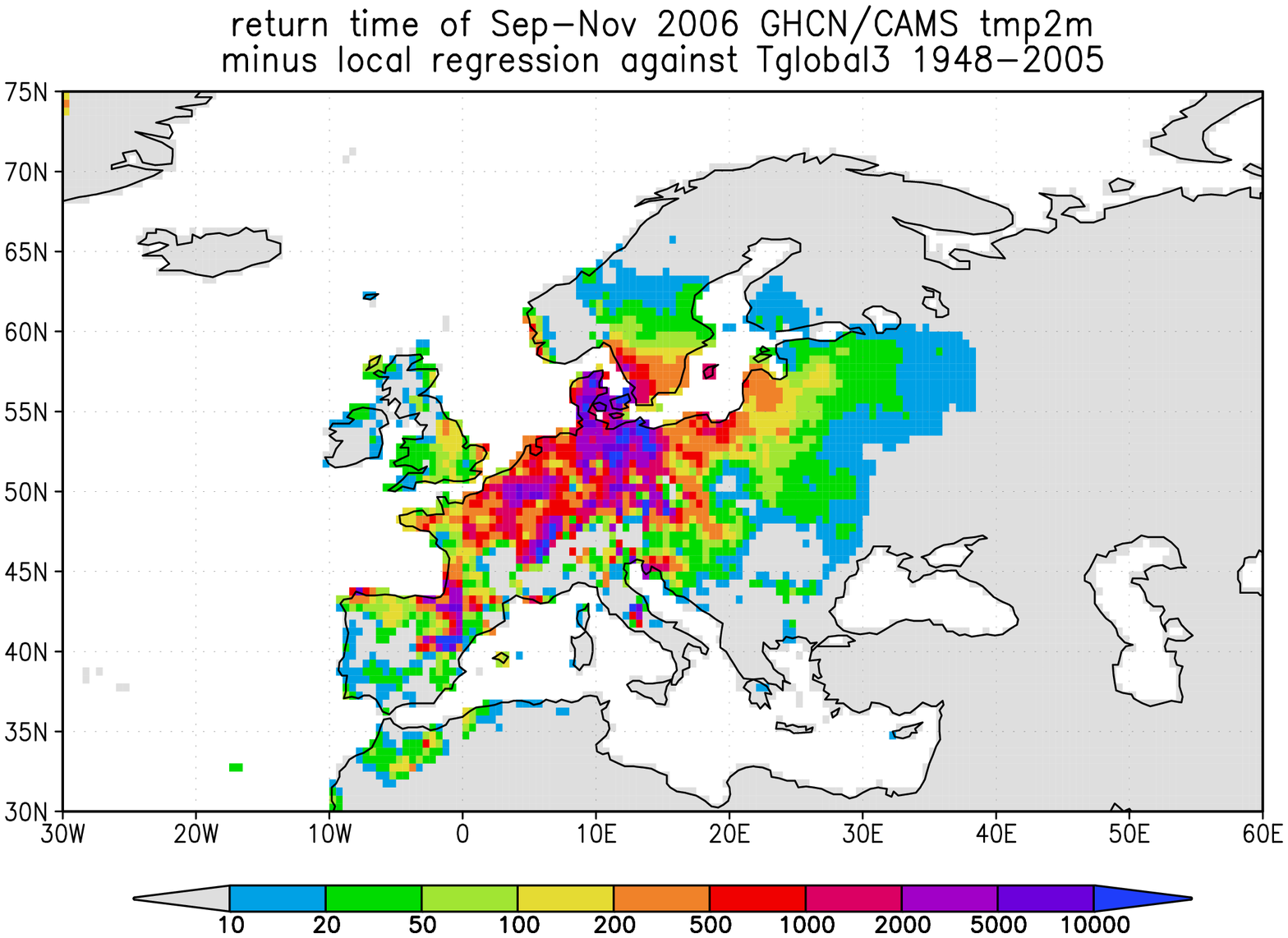}%
\hspace\columnsep%
\includegraphics[angle=-90,width=\columnwidth,clip]{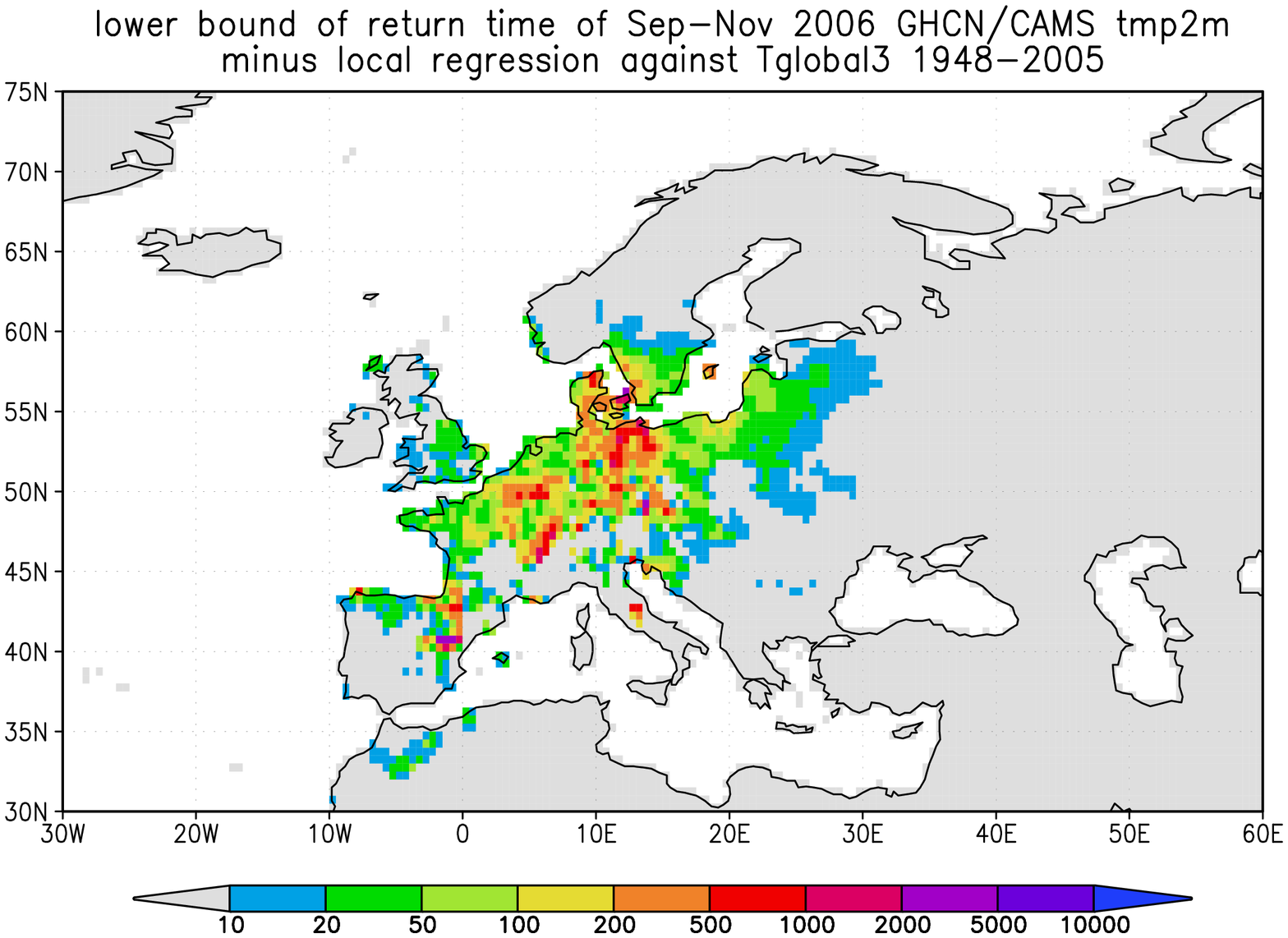}\\[-\baselineskip]
\makebox[\columnwidth][l]{\large a}%
\hspace\columnsep%
\makebox[\columnwidth][l]{\large b}%
\end{center}
\caption{As Fig.~\ref{fig:ECAgauss} but now with the local linear regression against the global mean temperature anomalies (with a 3-yr running mean, Fig.~\ref{fig:A}) subtracted (a), lower bound of the 95\% confidence interval on the return time (b).}
\label{fig:Tdebilt-Tglobal_eu}
\end{figure*}

The same extrapolation in the GHCN/CAMS dataset (Fig.~\ref{fig:Tdebilt-Tglobal_eu}) show the improbable weather to have extended over a large part of Europe, with estimates of return times of $\epsilon(t)$ longer than 200 years over most of the area where the anomaly was largest, reaching 2000 years in northern Germany.  The lower bounds of the two-sided 95\% confidence interval (determined as in Fig.~\ref{fig:Tdebilt-Tglobal_extra}) are more than 300 years there.  The Gaussian approximation probably underestimates the return times, as the anomalies $\epsilon(t)$ are negatively skewed in most of the area.

We conclude that global warming has made a temperature anomaly like the one observed in autumn 2006 between 10 and 50 times more likely than under the false assumption of stationary climate, with the larger factors near coasts where the trend is larger compared to natural variability.  Still, other factors than global warming conspired to give estimated return times well over 200 years in most of the area with large anomalies, reaching 2000 years (lower bound 300 years) in northern Germany.  A rare event indeed, even taking into account a shift in the probability density function proportional to global temperature changes, determined over the period before 2006.

\section{Circulation}

\begin{figure*}
\vspace*{2mm}
\begin{center}
\includegraphics[angle=-90,width=0.333\textwidth]{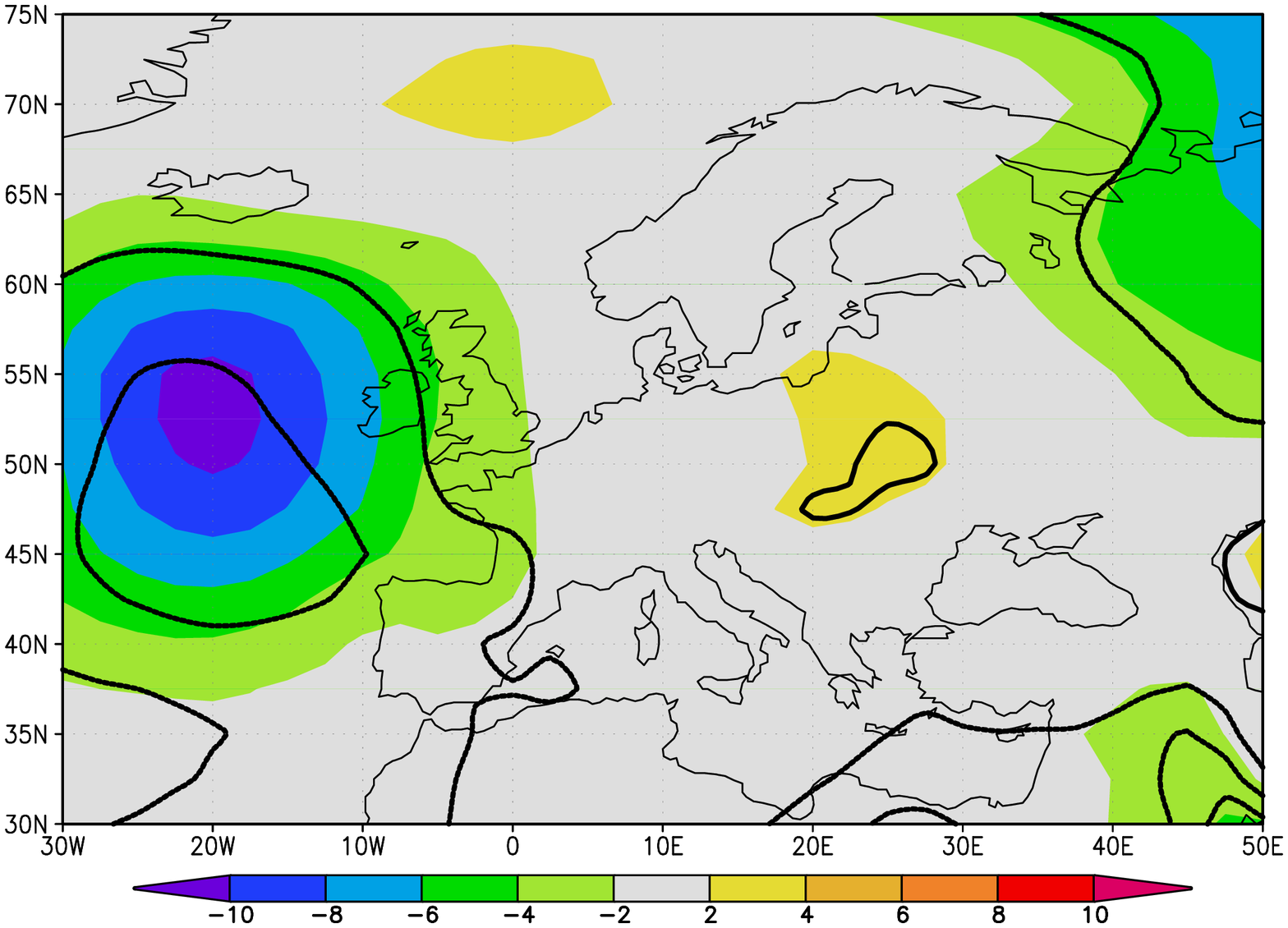}%
\includegraphics[angle=-90,width=0.333\textwidth]{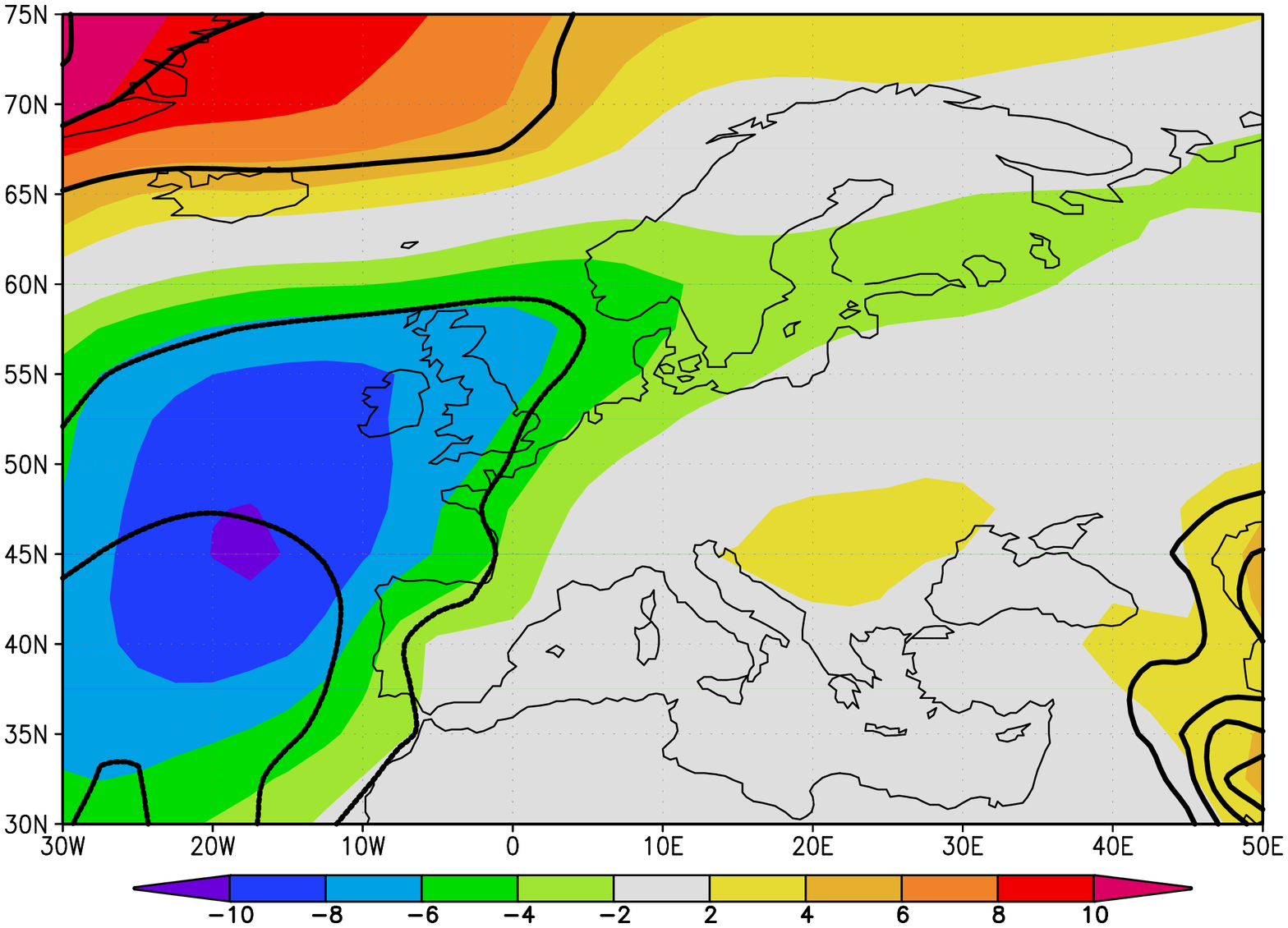}%
\includegraphics[angle=-90,width=0.333\textwidth]{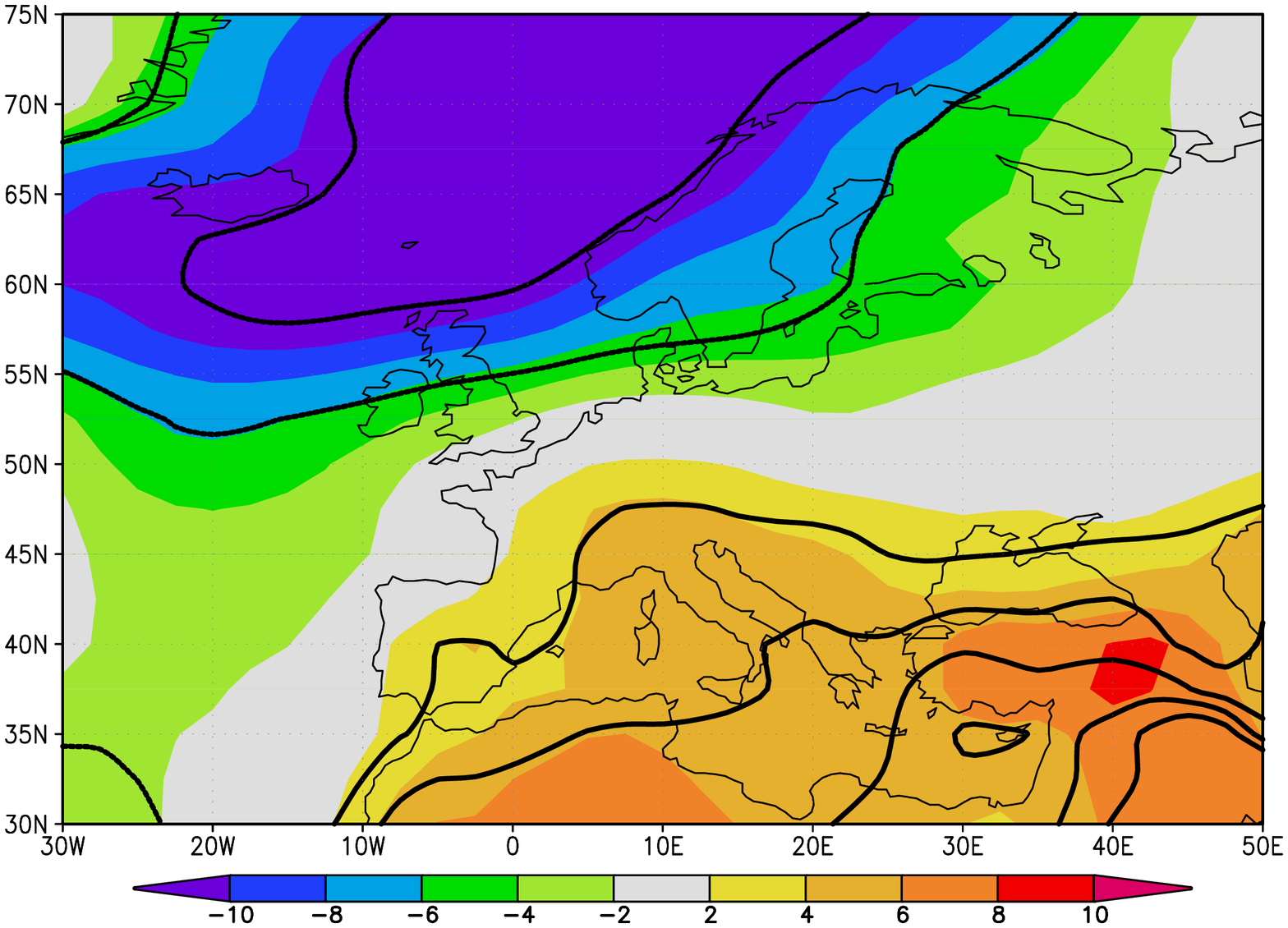}\\[-0.8\baselineskip]
\makebox[0.333\textwidth][l]{\large a}%
\makebox[0.333\textwidth][l]{\large b}%
\makebox[0.333\textwidth][l]{\large c}%
\end{center}
\caption{NCEP/NCAR sea-level pressure anomalies (hPa) relative to 1961-1990 in September (a), October (b) and November 2006 (c). The black lines indicate the number of standard deviations of the anomaly.}
\label{fig:slp}
\end{figure*}

It is obvious that a large part of the temperature anomalies in autumn 2006 was caused by the abnormal circulation in these months.
A persistent low-pressure area over the Atlantic Ocean caused predominantly south-easterly winds in September, southerly winds in October and south-westerly winds in November to the area north of the Alps (Fig.~\ref{fig:slp}).  In each of these months this corresponded to the direction advecting the highest temperatures.

The question how unusual the circulation patterns were is not easy to answer.  The local pressure anomalies did not reach $3\sigma$ deviations from the 1961-1990 normals (black lines in Fig.~\ref{fig:slp}).  A deviation of this size is not rare without restriction on the location.

In order to do a meaningful extreme value analysis on the circulation patterns these 2-dimensional fields have to be reduced to local time series.  One way to do this is to compute a circulation-dependent temperature.  In Europe, temperatures are determined to a large extend by wind direction due to large climatological temperature gradients.  Cloudiness also affects the temperature independently.  In the very simple model (VSM) of \citet{vanUldenvanOldenborgh2006} the anomaly in local monthly mean 2m temperature is therefore decomposed as
\begin{eqnarray}
\label{eq:vsm1}
T'(t) & = & T'_\mathrm{circ} + T'_\mathrm{noncirc}(t) + M T'(t-1)\\
T'_\mathrm{circ} & = & A_W G'_\mathrm{west}(t)\label{eq:Tcirc}
+ A_S G'_\mathrm{south}(t) + B V'(t)\\
T'_\mathrm{noncirc}(t) & = & A_g T^{\prime(3)}_\mathrm{global}(t) + \eta(t)\,.
\label{eq:vsm3}
\end{eqnarray}
The effect of circulation on the temperature is approximated by the circulation-dependent temperature anomalies $T'_\mathrm{circ}$, which are linearly proportional to the zonal and meridional geostrophic wind anomalies $G'_\mathrm{west}$ and $G'_\mathrm{south}$ and a measure for cloudiness, the vorticity $V'$.  The geostrophic wind is computed with the corners of a 20\dg\ longitude by 10\dg latitude box; the vorticity as the difference between the average of the corners and the central value.  (All anomalies are relative to the mean observed values for 1961--1990.)

The non-circulation-dependent temperature anomalies $T'_\mathrm{noncirc}(t)$ consist of the part linearly proportional to global warming and the remaining noise $\eta(t)$.  Finally, $M$ is a memory term for past local temperature. This term is modelled as a regression on the previous months' anomalous temperature \citep{vandenDoolNap1981}.
The geostrophic winds and vorticity are computed from the NCEP/NCAR reanalysis sea-level pressure \citep{Kalnay1996}.

There is some ambiguity in this model if the climate change involves changes in the circulation patterns parametrised by the geostrophic wind.  However, the interannual variations in circulation have so far been much larger than the long-term shifts, so that in practice the terms $A_S,A_W,B$ are most strongly influenced by the interannual variability and this part of climate change is not contained in the circulation-independent temperature changes.  In autumn, there has been no 90\% significant change in sea-level pressures in Europe apart from a slight increase over the Balkan.

\begin{figure*}
\vspace*{2mm}
\begin{center}
\includegraphics[angle=-90,width=\columnwidth]{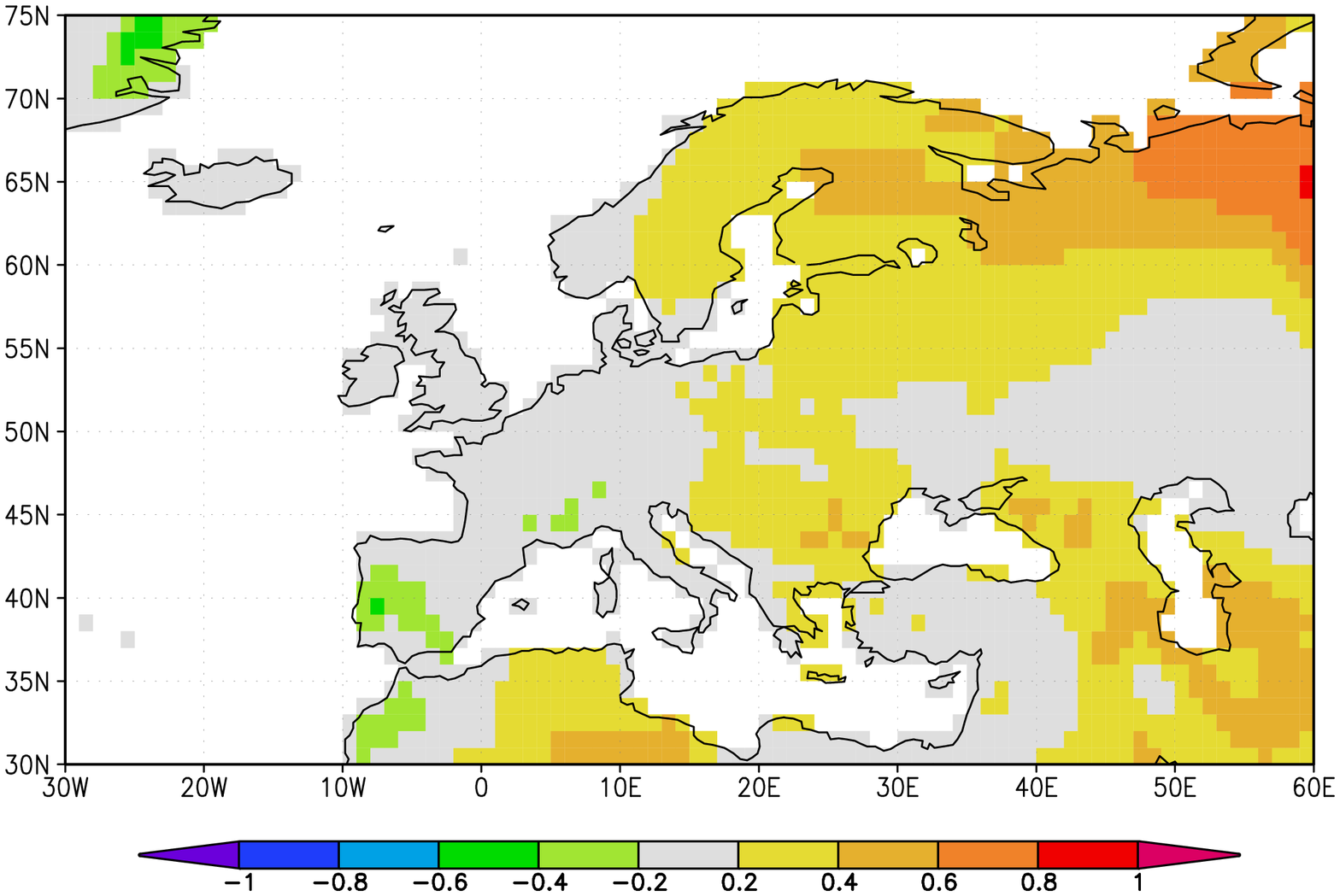}%
\hspace\columnsep%
\includegraphics[angle=-90,width=\columnwidth]{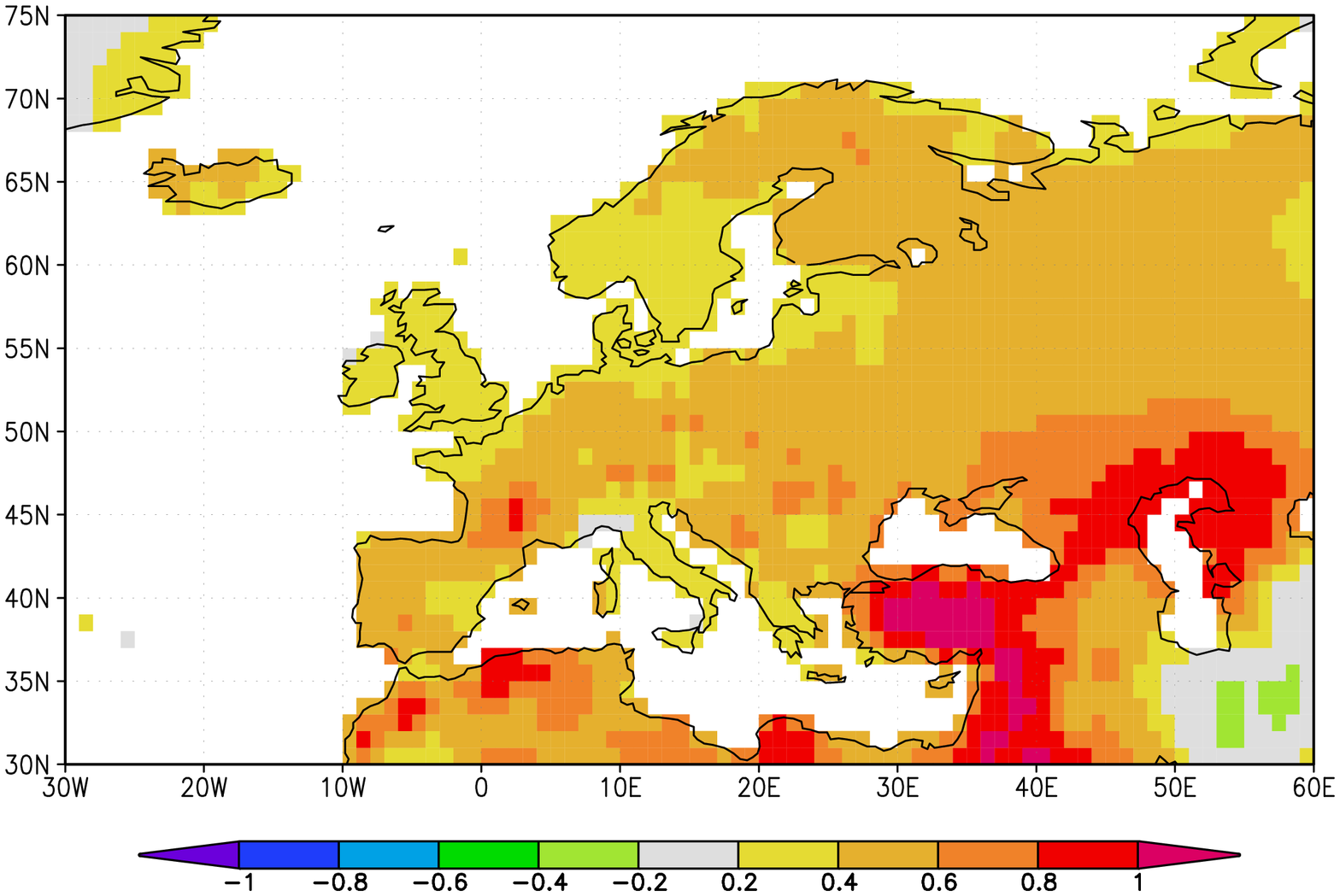}\\[-1.4\baselineskip]
\makebox[\columnwidth][l]{\large a}%
\hspace\columnsep%
\makebox[\columnwidth][l]{\large b}\\
\includegraphics[angle=-90,width=\columnwidth]{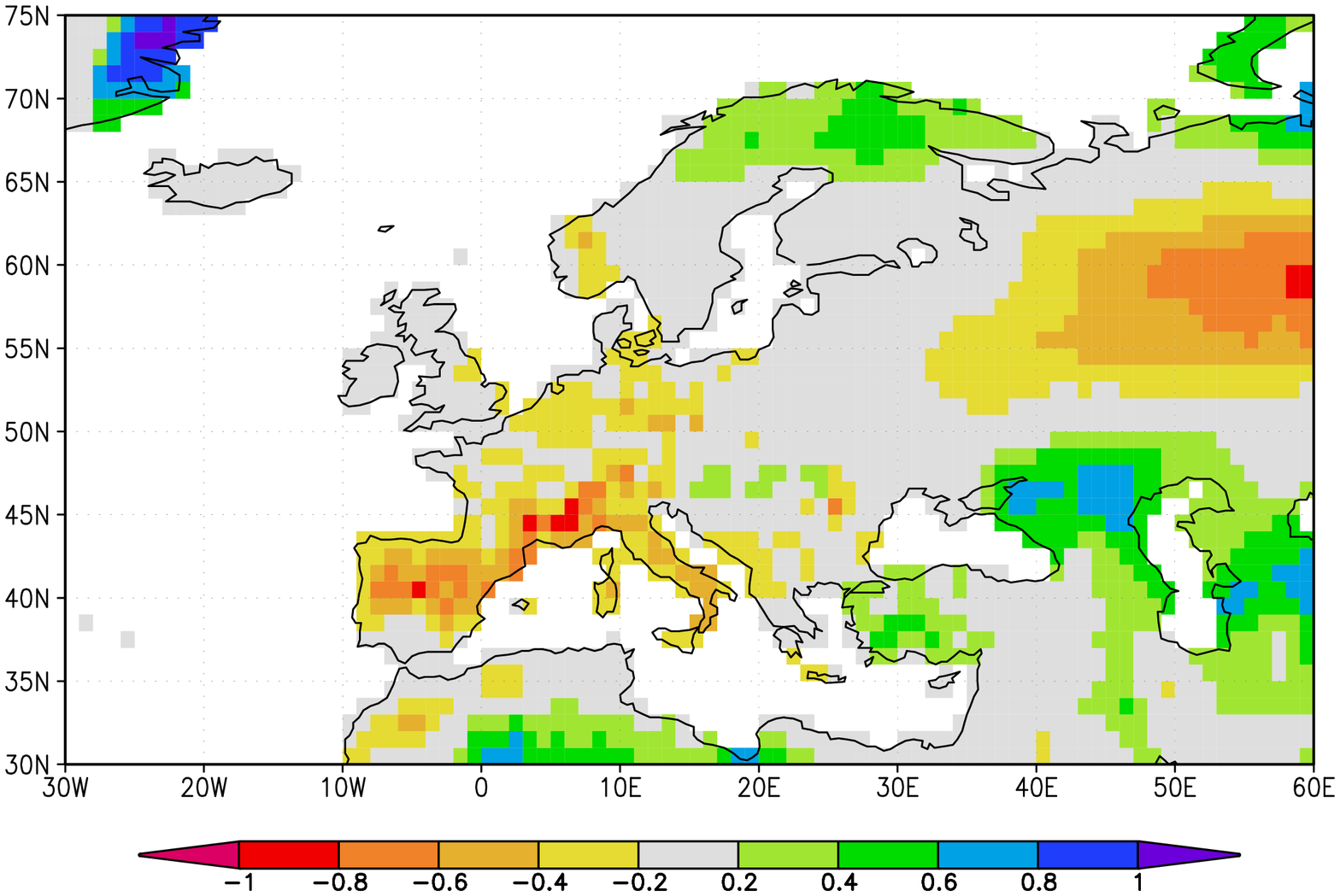}%
\hspace\columnsep%
\includegraphics[angle=-90,width=\columnwidth]{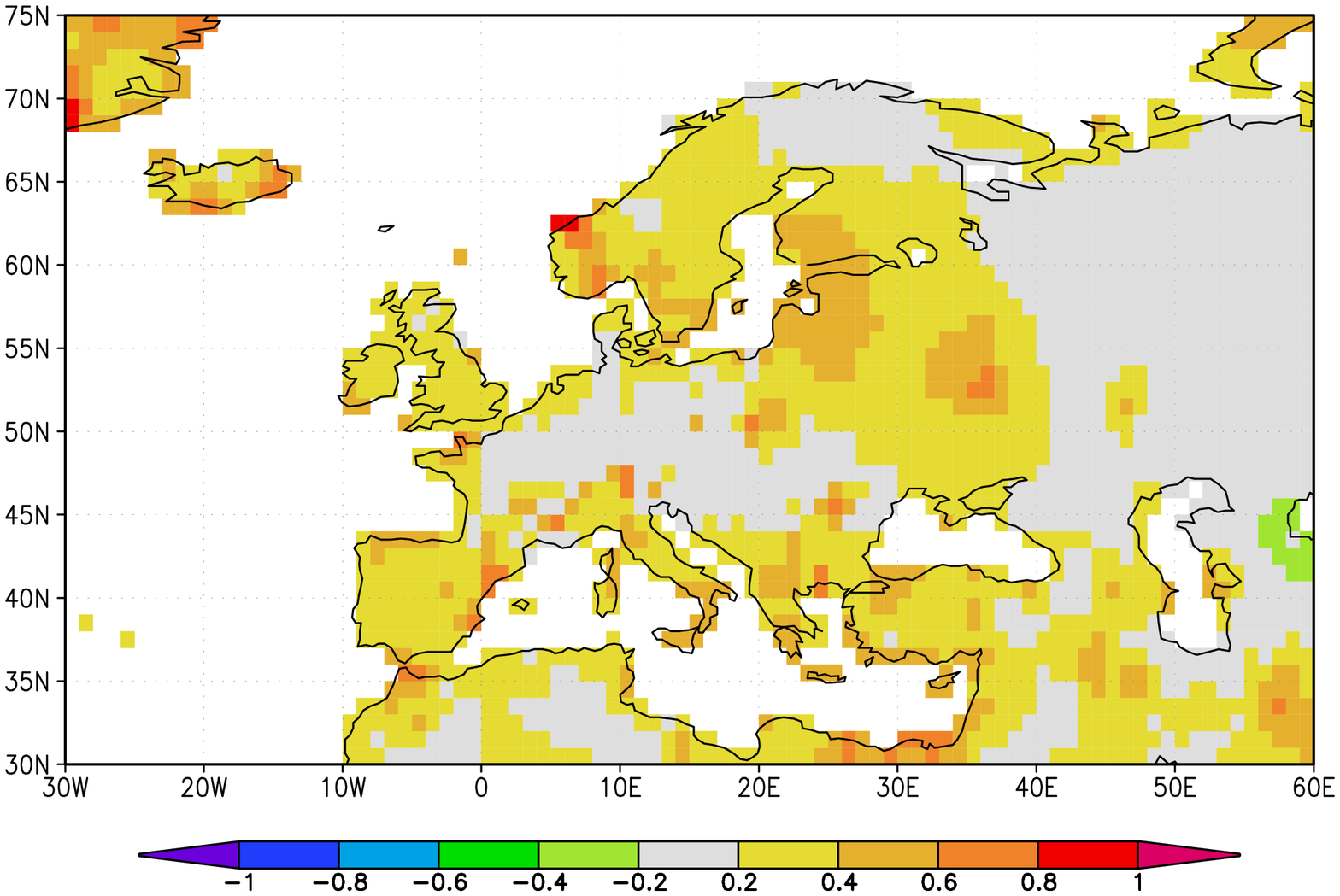}\\[-1.4\baselineskip]
\makebox[\columnwidth][l]{\large c}%
\hspace\columnsep%
\makebox[\columnwidth][l]{\large d}
\end{center}
\caption{The coefficients of the VSM Eqs~(\ref{eq:vsm1}-\ref{eq:vsm3}) averaged over September-November: zonal geostrophic wind $A_W [\mathrm{Km}^{-1}\mathrm{s}]$ (a), the meridional zonal wind $A_S [\mathrm{Km}^{-1}\mathrm{s}]$ (b), the vorticity $B [\mathrm{K\:hPa}^{-1}]$ (c); the memory term $M$ [1] in September (d).}
\label{fig:coefs}
\end{figure*}

The five model coefficients are fitted at each grid point over 1948-2005 (averaged to $1\dg\times1\dg$ for computational reasons).  Averaged over the autumn, the coefficients $A_W$ and $A_S$ reflect the gradients in the climatological temperature over Europe  (Fig.~\ref{fig:coefs}).  In this season the southerly component is most important in determining the temperature.  Positive vorticity leads to more sunshine, which has a positive influence on temperature in central and southern Europe, but in northern Europe a lack of clouds increases night-time radiation more and hence cools the surface.  The memory term is large near seas due to the thermal inertia of sea water on the monthly time scale.

\begin{figure}
\vspace*{2mm}
\begin{center}
\includegraphics[angle=-90,width=\columnwidth]{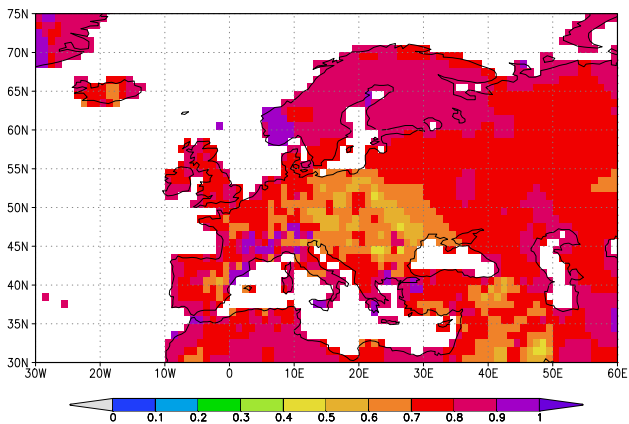}%
\end{center}
\caption{Correlation of the September-November temperature anomaly from the VSM with $\eta(t)=0$ and the observed temperature 1948-2005.}
\label{fig:corr}
\end{figure}

The VSM Eqs~(\ref{eq:vsm1}--\ref{eq:vsm3}) explains over half the variance of the temperature, i.e., the correlation between the modelled  temperature with $\eta(t) = 0$ and the observed temperature is about $r=0.7$ to $0.8$ in autumn in Europe (Fig~\ref{fig:corr}).

\begin{figure*}
\vspace*{2mm}
\begin{center}
\includegraphics[angle=-90,width=\columnwidth]{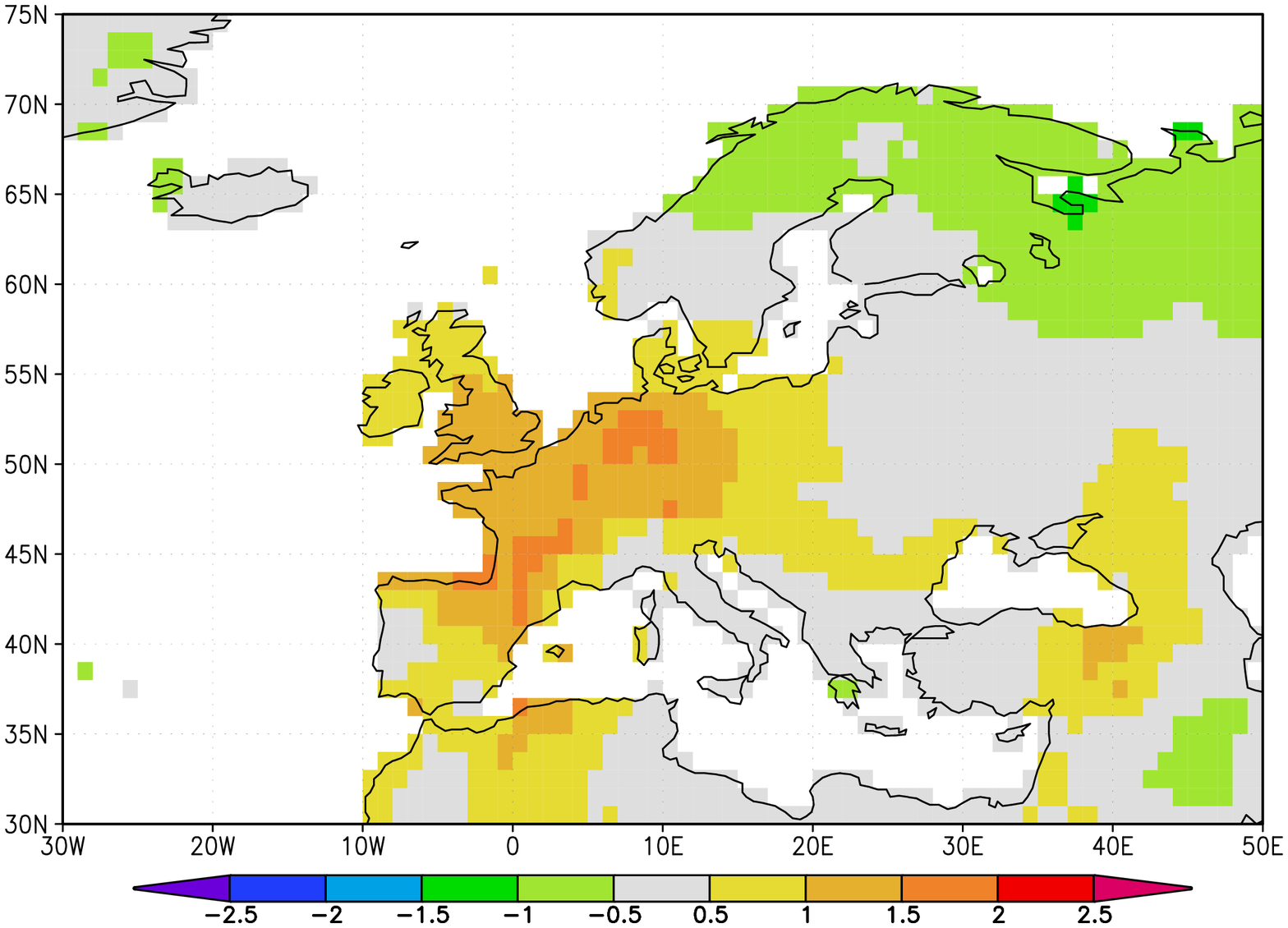}%
\hspace\columnsep%
\includegraphics[angle=-90,width=\columnwidth]{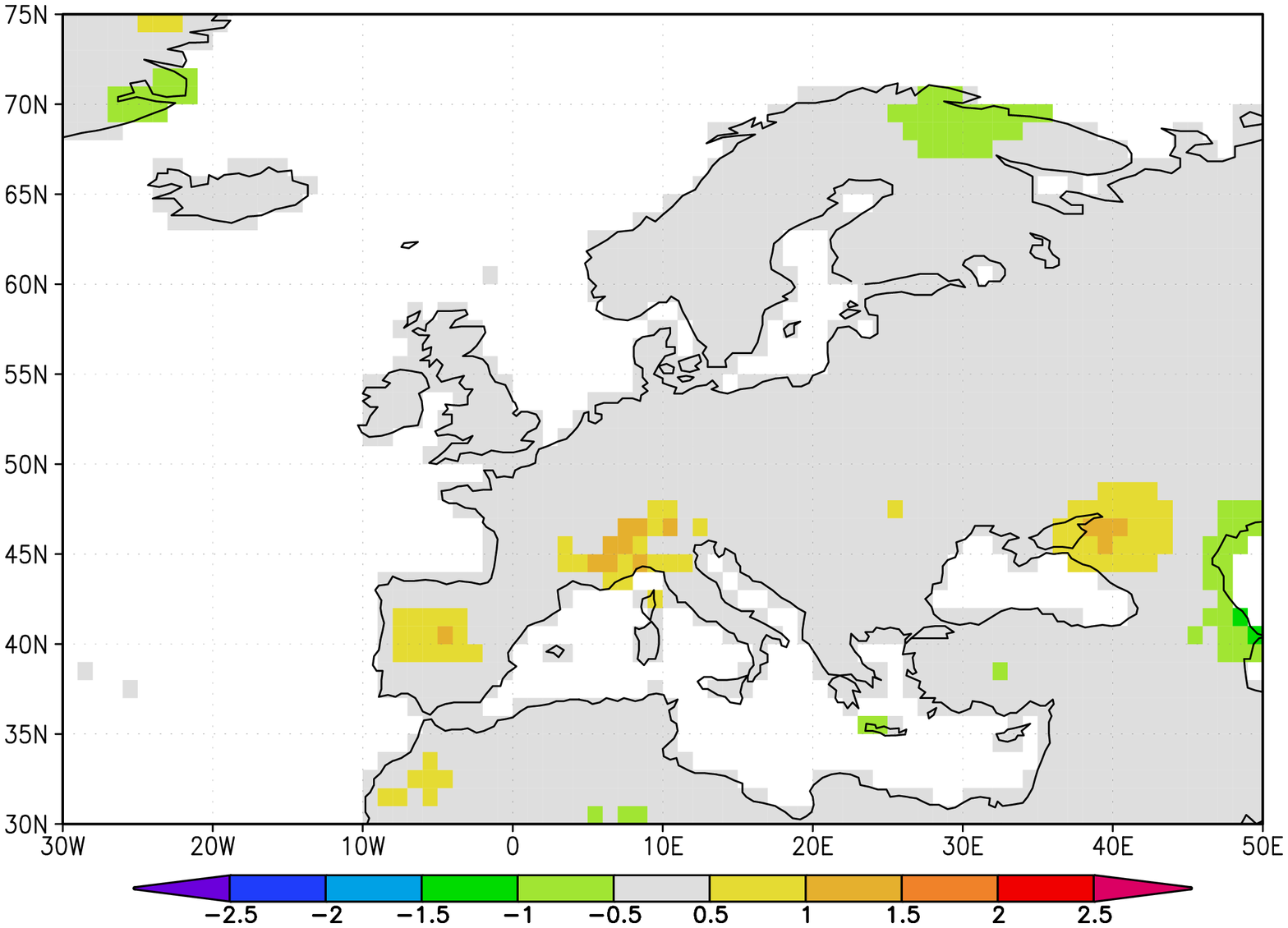}\\[-1.4\baselineskip]
\makebox[\columnwidth][l]{\large a}%
\hspace\columnsep%
\makebox[\columnwidth][l]{\large b}
\end{center}
\caption{Contribution (degrees Celsius) of the geostrophic wind term (a) and vorticity term (b) to the temperature anomaly in autumn 2006.}
\label{fig:factors}
\end{figure*}

In Fig.~\ref{fig:factors} the contribution of the various terms for the 2006 event in the VSM are shown.  The anomalous circulation contributed 1.5--2.0\dg C to the observed anomaly within the linear framework of Eqs~(\ref{eq:vsm1}-\ref{eq:vsm3}).  The anomalous vorticity that gave rise to large amount of sunshine in September increased the temperature by less than 0.5\dg C in the seasonal average in this linear approximation.

On the shores of the North Sea persistence has also contributed.  However, this was not due to the below-normal temperatures in August.  The North Sea was still warm from the exceptionally high temperatures in July.  This is not captured by the VSM, which only depends on the previous month, hence we cannot give a quantitative estimate.  Based on the observed SST anomalies of about 2\dg C at the beginning of September it is estimated that this contributed roughly half a degree to the autumn temperature anomaly in De Bilt.

\begin{figure}
\vspace*{2mm}
\begin{center}
\includegraphics[angle=-90,width=\columnwidth,clip]{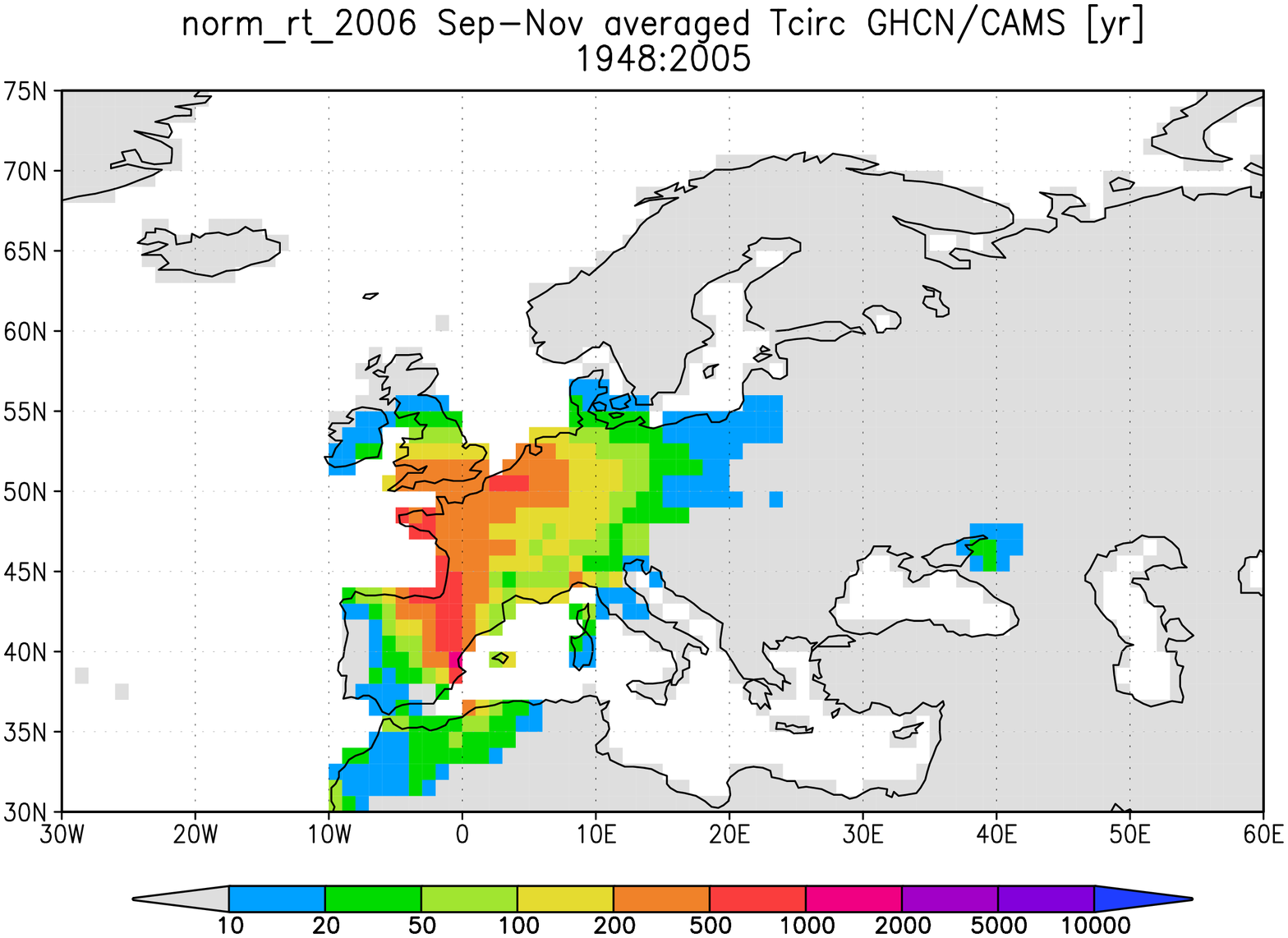}%
\end{center}
\caption{As Fig.~\ref{fig:Tdebilt_extra} but only for the circulation-dependent temperature $T_\mathrm{circ}$ defined by Eq.~\ref{eq:Tcirc}.}
\label{fig:norm_rt_tcirc}
\end{figure}

Finally, the return times of $T_\mathrm{circ}$ in autumn 2006 are shown in Fig.~\ref{fig:norm_rt_tcirc}.  Comparing this with Fig.~\ref{fig:Tdebilt-Tglobal_eu}, the linear effect of the circulation in 2006 explains most of the return times in the western half of the area with high return times, but in Germany other effects were responsible for the high anomalies.  As the memory term is also small in this area, these are not described by the simple model.

\section{Climate model simulations}

Observed autumn temperatures are far out of the range observed so far, even after shifting the distribution with an amount proportional to global mean temperature anomalies.   Do current climate models predict this type of events to happen more frequently as Europe heats up?  There are two caveats when using these models:
\begin{enumerate}
\item the midlatitude circulation response of climate models varies greatly from model to model and up to now lacks a sound theoretical footing \citep{vanUldenvanOldenborgh2006,Miller2006};
\item the local temperature response to global warming is uncertain. \end{enumerate}

To investigate the changes in the distribution of autumn temperatures results from the 17 standard runs with the ECHAM5/MPI-OM1 model \citep{MPI-ECHAM5} runs of the ESSENCE project\footnote{See www.knmi.nl/$\sim$sterl/Essence/ for details} were used. These cover the period 1950-2100 using observed concentrations of greenhouse gases and aerosols up to 2000 and the SRES A1B scenario afterwards.  This model simulates the mean circulation over Europe reasonable well on monthly time scales \citep{vanUldenvanOldenborgh2006,KNMIscenarios}.  In simulations of the 20th century (20c3m experiments) the monthly mean sea-level pressure fields resemble those of the observations best of the models in the World Climate Research Programme's (WCRP's) Coupled Model Intercomparison Project phase 3 (CMIP3) multi-model dataset.

An estimate of the systematic errors is provided by a comparison with other models that simulate a reasonable mean climate over Europe in this measure.  These are GFDL CM2.1 \citep{GFDL-CM2.0}, MIROC 3.2 T106 \citep{MIROC3.2}, HadGEM1 \citep{UKMO-HadGEM1} and CCCMA CGCM 3.2 T63 \citep{CGCM3.1}.  The MIROC high resolution model did not have enough data to reconstruct changes in the full temperature distribution, only the mean.

The ECHAM5/MPI-OM model used in ESSENCE simulates the global mean temperature very well, with a ratio between observed and modelled trends of $1.10\pm0.07$ ($1\sigma$ errors).  The GFDL CM2.1 model has similar agreement, but the other models overestimate the trend in the global mean temperature over 1950-2006 by factors 1.5 (HadGEM1), 1.6 (MIROC) and 2.0 (CCCMA).  To account for these biases, we defined the local trend as a regression against modelled global mean temperature, as was done in \citet{KNMIscenarios}.   The local temperature rise as a function of time rather than global mean temperature is higher by the same factors 1.5 to 2.0 in these models.

\begin{figure*}
\vspace*{2mm}
\begin{center}
\raisebox{32mm}{%
\makebox[0pt][l]{\large a}%
\includegraphics[angle=-90,width=\columnwidth]{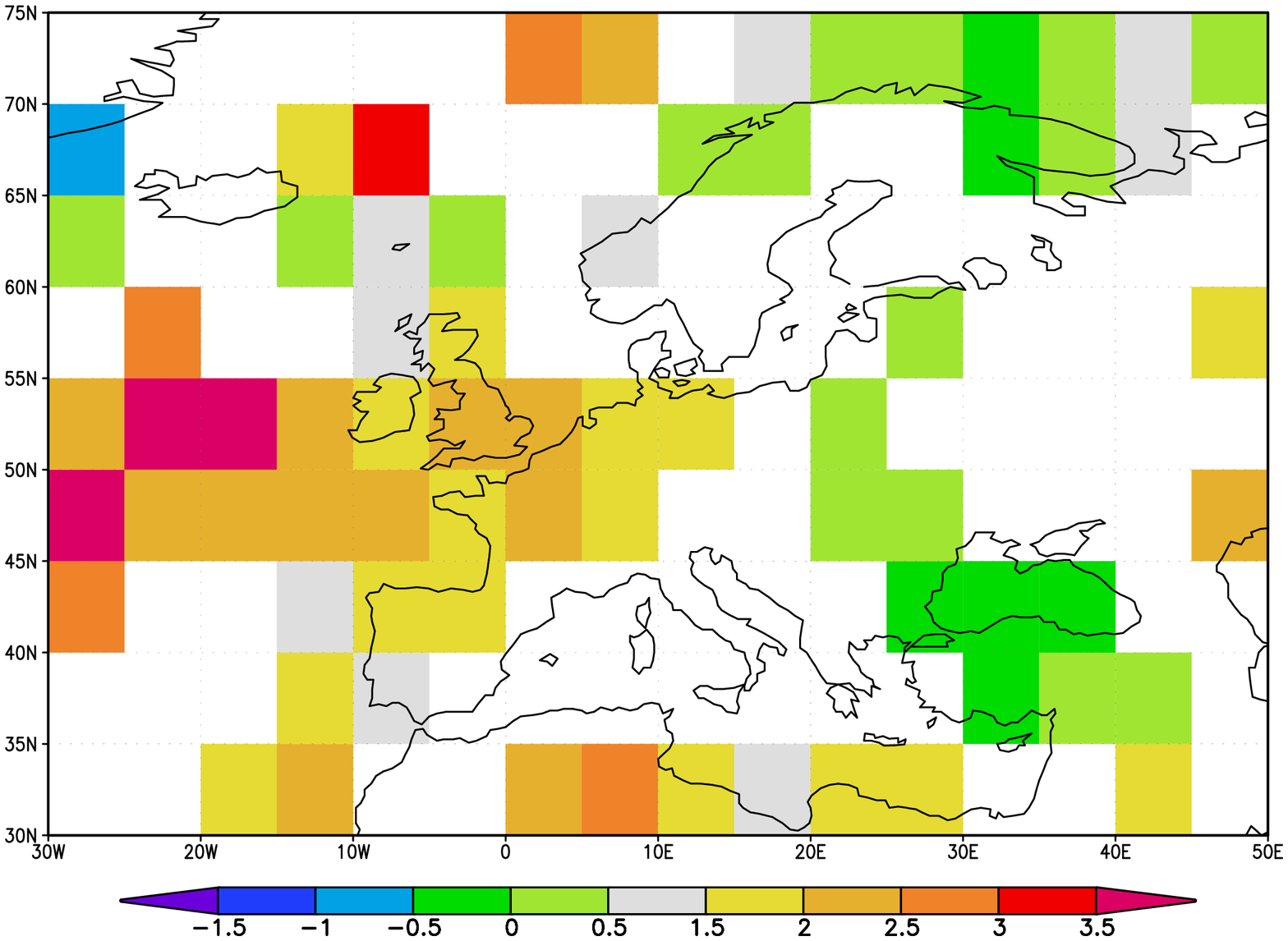}}%
\hspace\columnsep%
\shortstack{%
\makebox[0pt][l]{\large b}%
\includegraphics[angle=-90,width=0.5\columnwidth]{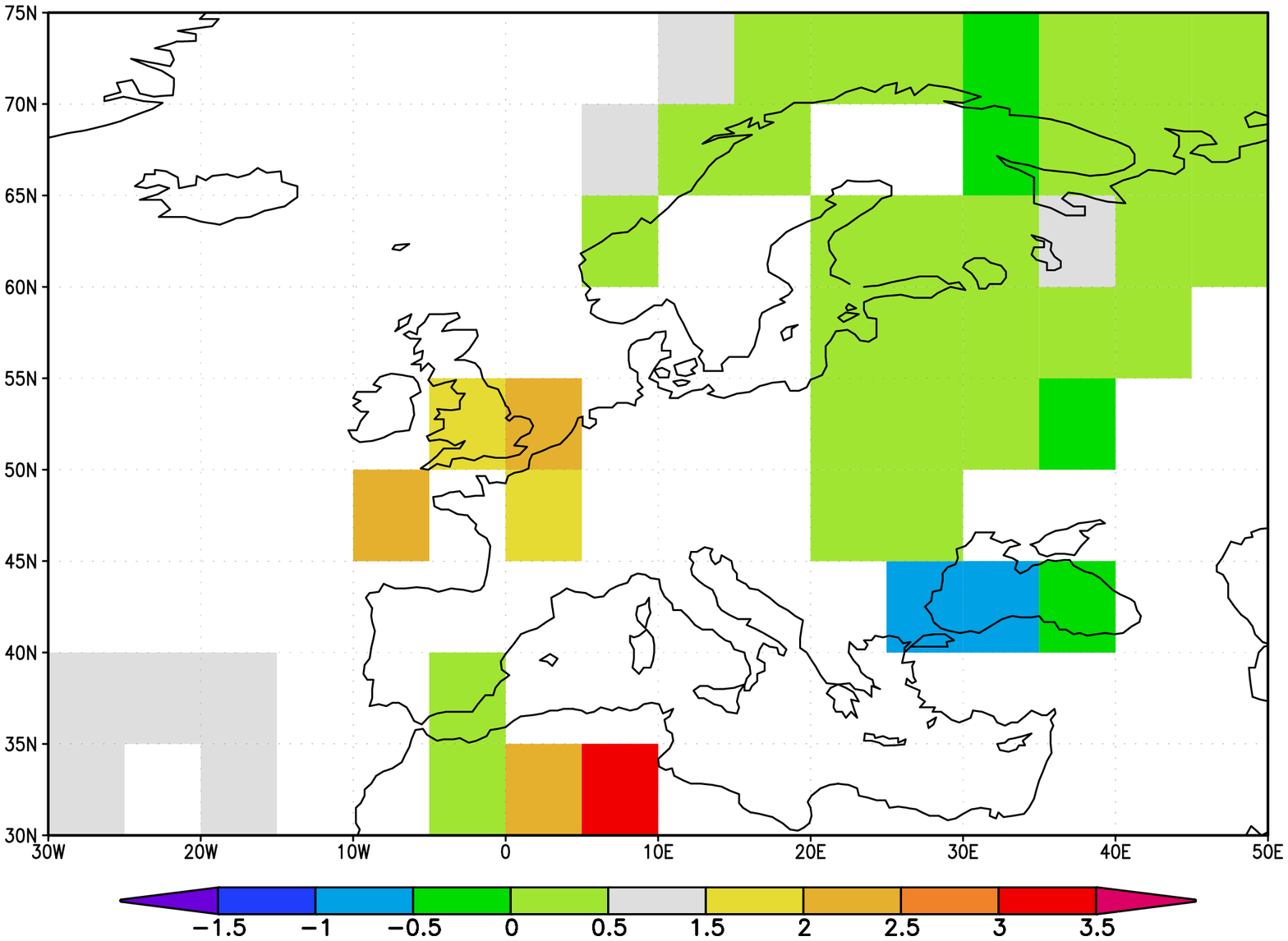}%
\makebox[0pt][l]{\large c}%
\includegraphics[angle=-90,width=0.5\columnwidth]{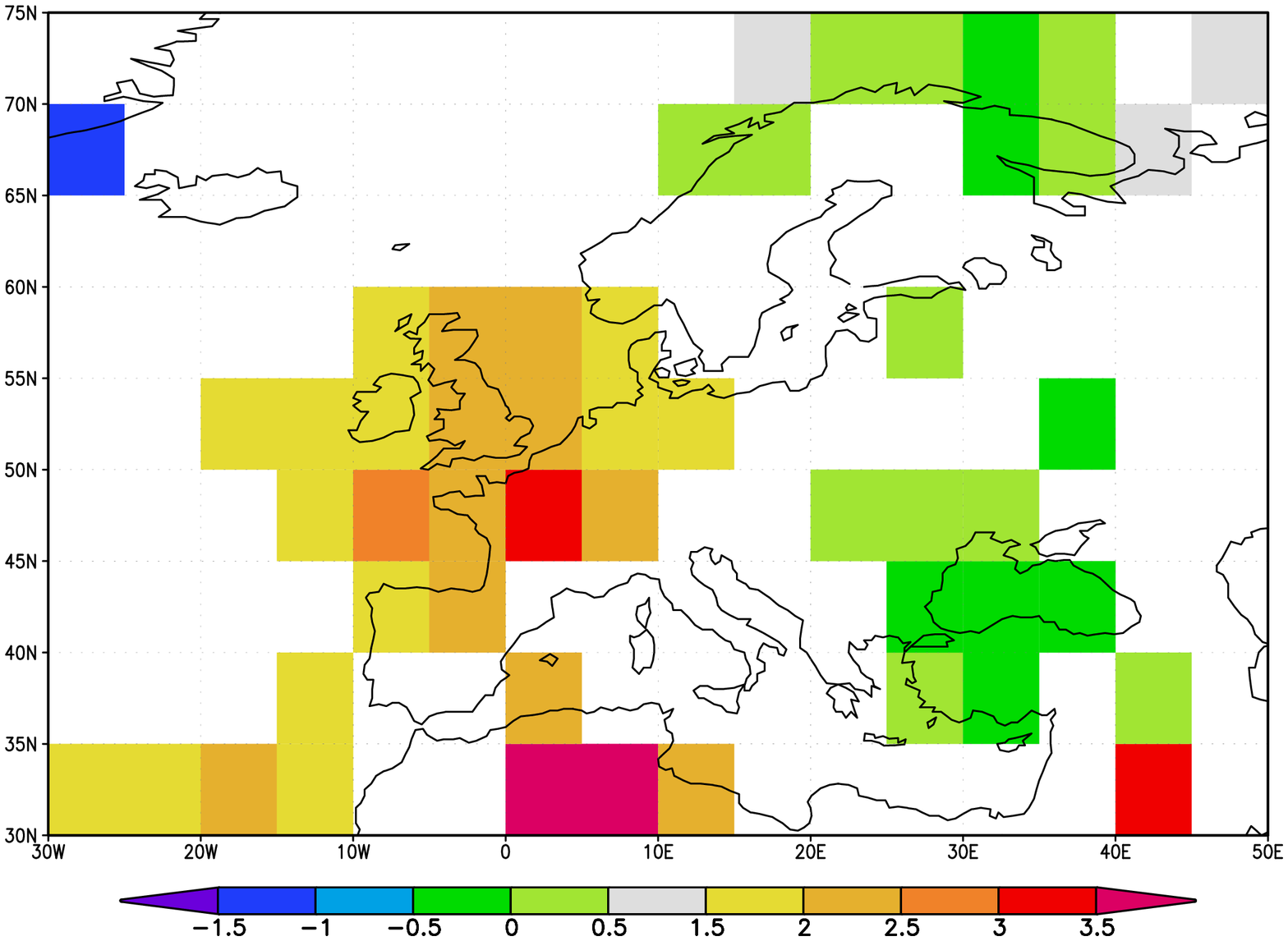}\\
\makebox[0pt][l]{\large d}%
\includegraphics[angle=-90,width=0.5\columnwidth]{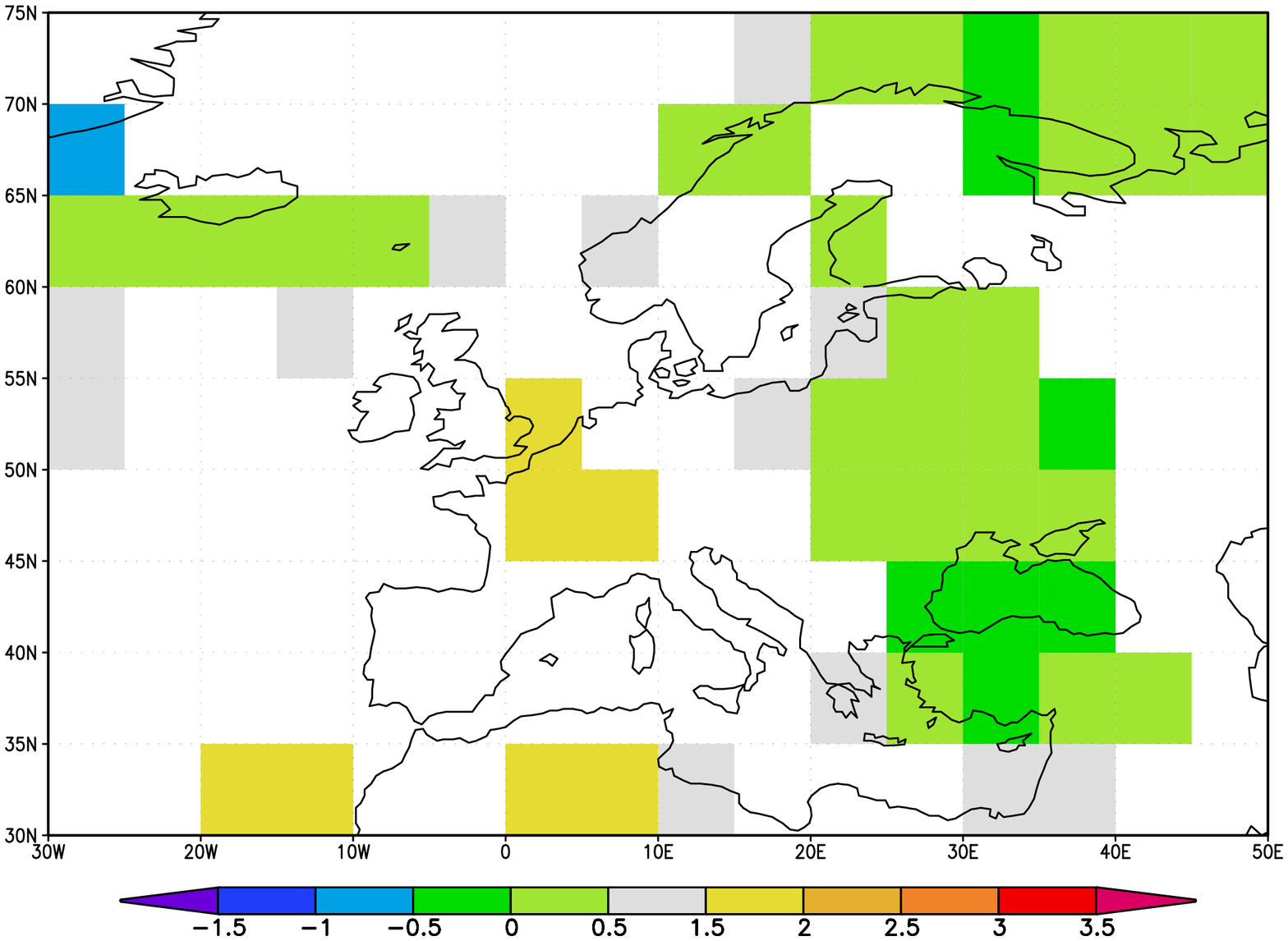}%
\makebox[0pt][l]{\large e}%
\includegraphics[angle=-90,width=0.5\columnwidth]{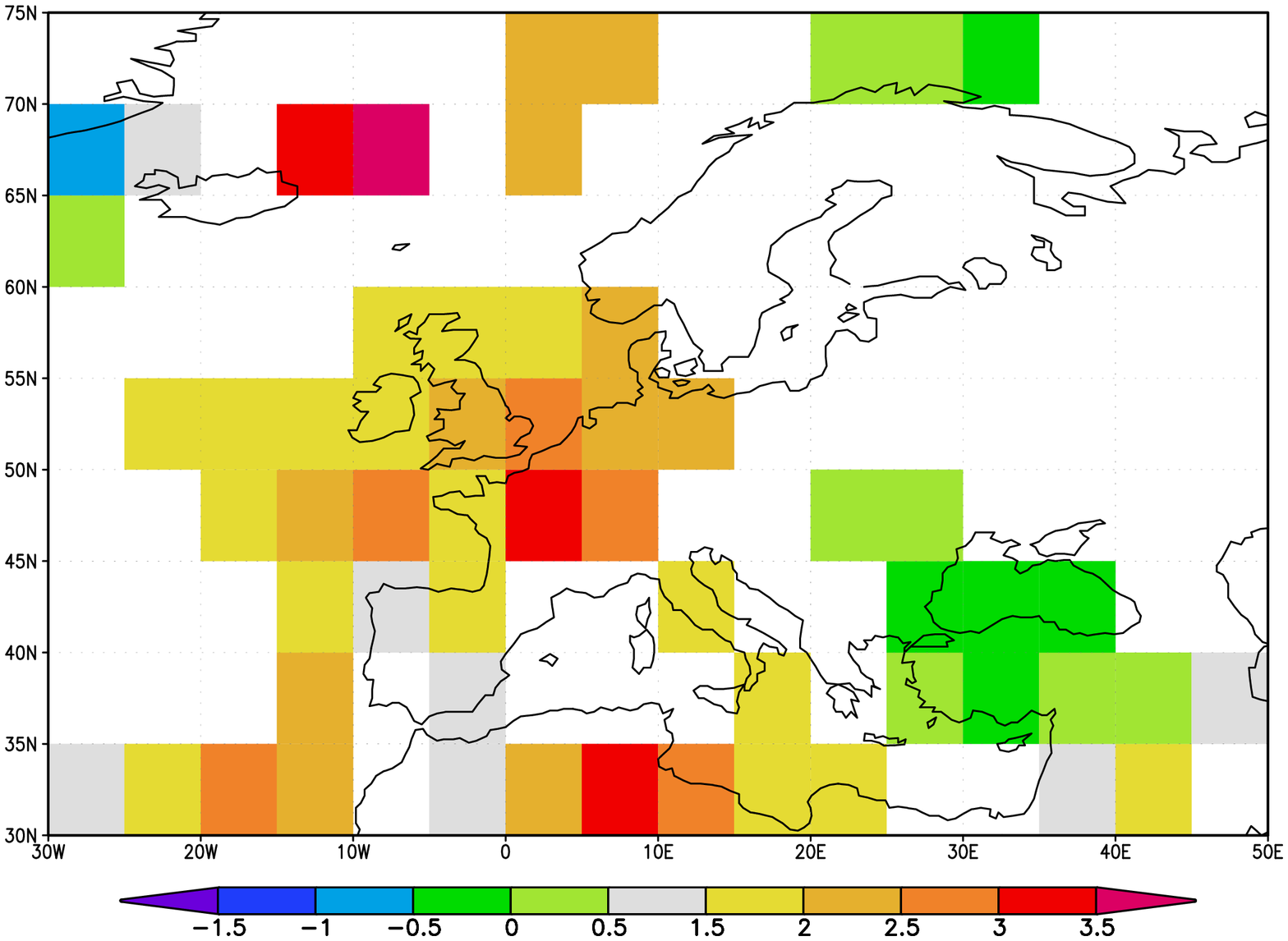}}%
\end{center}
\caption{Ratio of observed and modelled trends 1950-2006.  ESSENCE (ECHAM5/MPI-OM) (a), GFDL CM2.1 (b), MIROC 3.2 T106 (c), UKMO HadGEM1 (d) and CCCMA CGCM3.1 T63 (e).  Only grid points where the difference with one is at least one standard error (assuming a normal distribution) are shown.  The model trends have been computed as a regression against the modelled global mean temperature.}
\label{fig:ratio}
\end{figure*}

Figure~\ref{fig:ratio} shows the ratio of observed and modelled warming trends in Europe in autumn.  The ECHAM5/MPI-OM model is seen to underestimate the warming trend in the area of the observed extreme by a factor 1.5 or more.  In the other models the ratio between local observed and modelled temperature trends has larger errors as there are fewer ensemble members available, but these models also show a higher observed than modelled warming relative to the rest of the world in the areas of the autumn 2006 anomaly.

\begin{figure}
\vspace*{2mm}
\begin{center}
\includegraphics[width=\columnwidth,clip]{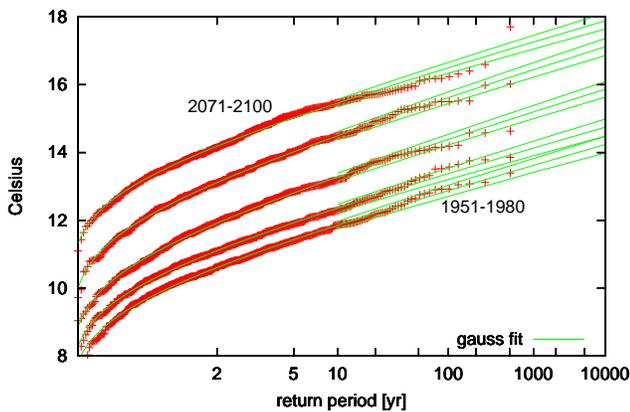}
\end{center}
\caption{Extreme autumn temperatures at 52\dg N, 5\dg E in 17 ECHAM5/MPI-OM1 20C3M/SRES A1B runs in 1951-1980, 1981-2010, 2011-2041, 2041-2070 and 2070-2100.}
\label{eq:Tessence_extra}
\end{figure}

Figure~\ref{eq:Tessence_extra} shows the extreme value distribution of the model surface air temperature at the position of De Bilt for different 30-year intervals.  The distributions are described well by a Gaussian.  Above the linear increase in temperature proportional to the global mean temperature rise, there is no indication of any change in the distribution that would make extremely warm autumn temperatures more likely.  In summer this change to more positive skewness is clearly seen by steeper slopes in the cumulative distributions (not shown); this can be understand from soil moisture effects \cite[e.g.,][]{SchaerJendritsky2004,Seneviratne2006,Fischer2007}.

\begin{figure*}
\vspace*{2mm}
\begin{center}
\makebox[0pt][l]{\includegraphics[width=0.333\textwidth]{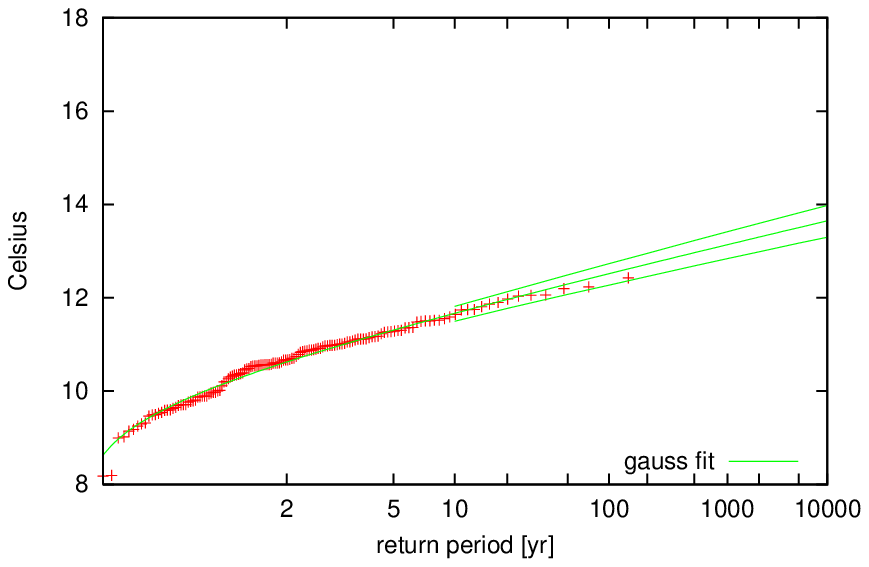}}%
\includegraphics[width=0.333\textwidth]{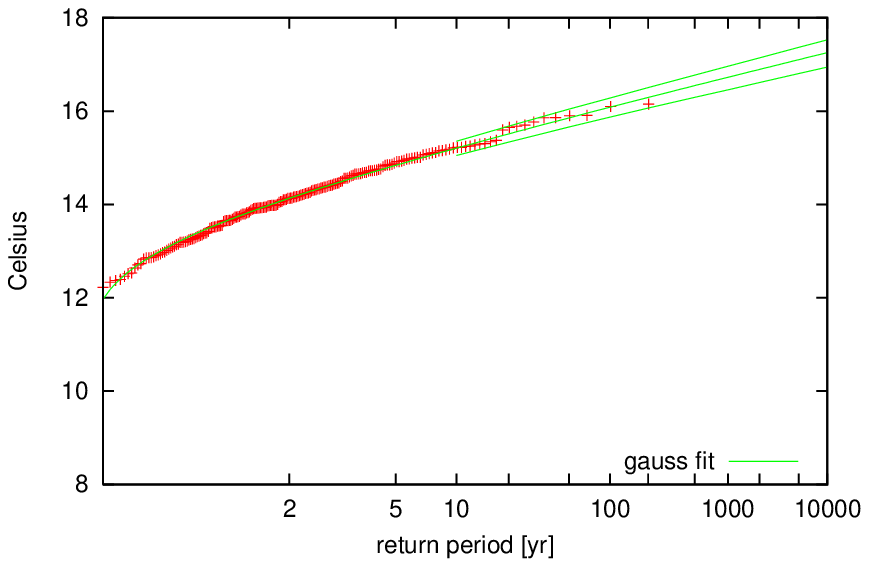}%
\makebox[0pt][l]{\includegraphics[width=0.333\textwidth]{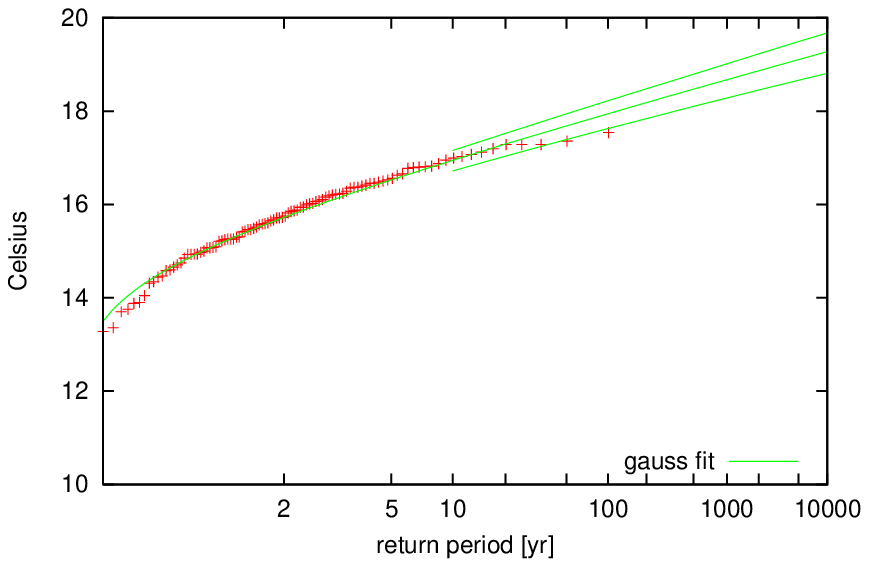}}%
\includegraphics[width=0.333\textwidth]{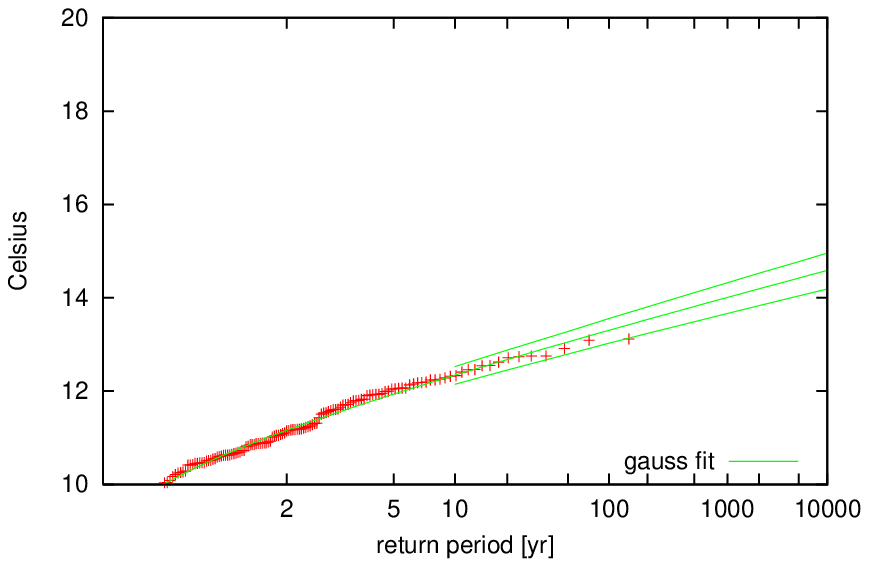}%
\makebox[0pt][l]{\includegraphics[width=0.333\textwidth]{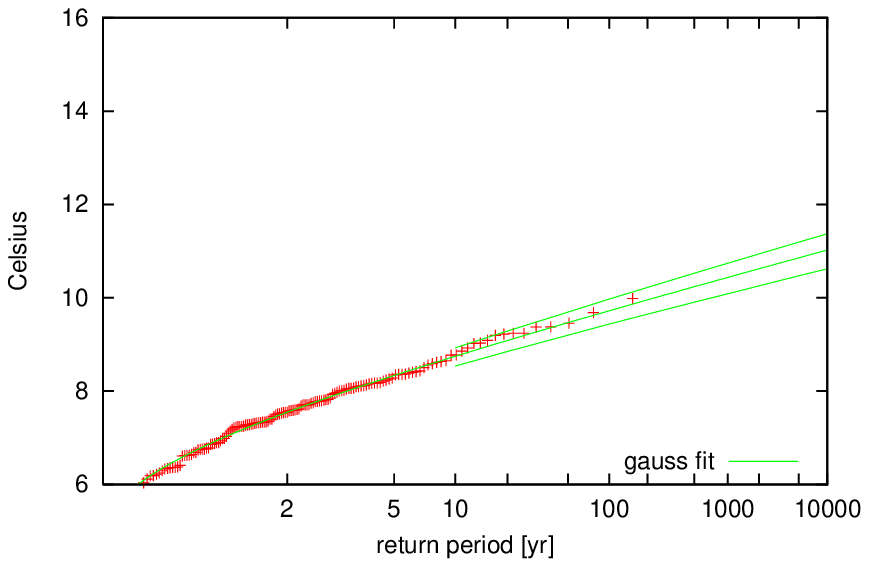}}%
\includegraphics[width=0.333\textwidth]{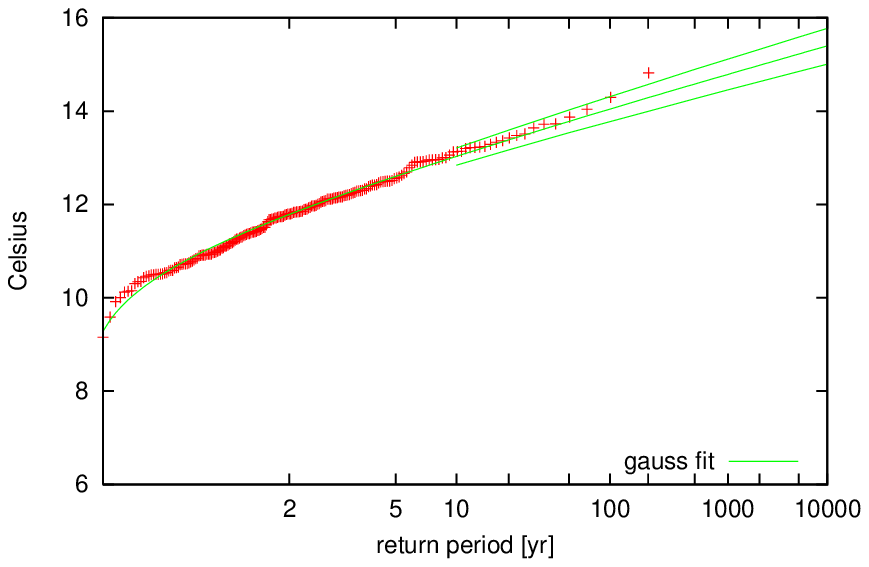}\\[-\baselineskip]
\makebox[0.333\textwidth][l]{\large \quad a}%
\makebox[0.333\textwidth][l]{\large \quad b}%
\makebox[0.333\textwidth][l]{\large \quad c}%
\end{center}
\caption{Extreme autumn temperatures at 52\dg N, 5\dg E in the 20C3M and SRES A1B stabilisation runs in GFDL CM2.1 (a), UKMO HadGEM1 (b) and CCCMA CGCM 3.2 T63 (c).}
\label{fig:Tmodels_extra}
\end{figure*}

This result was confirmed for the other climate models with a reasonable circulation over Europe for which a comparison between the 22nd and 23rd century with the 20th century could be made.  None of these show an increase in the slope of the cumulative distribution at the grid point corresponding to De~Bilt, Fig.~\ref{fig:Tmodels_extra}.

The probability of temperatures as high as those observed in 2006 increases if global warming shifts the distribution to higher values, or the width of the distribution increases towards higher temperatures.  The climate models considered here show no evidence for either of these mechanisms: they simulate a lower shift of the distribution and no change in shape.  Note that the change in width is impossible to obtain from observations.

\conclusions
The autumn of 2006 was extraordinarily warm in large parts of Europe, with temperatures up to 4\dg C above the 1961-1990 normals.  Assuming an unchanging climate, this would correspond to return times of 10000 years and more.

Global warming has made a warm autumn like the one observed in 2006 much more likely by shifting the temperature distribution to higher values.  Taking only this mean warming into account, the best estimate of the return time of the observed temperatures in 2006 still is more than 200 years in large parts of Europe.  The lower bound is 100 years or more years, reached when the trend is much larger than estimated from the years before 2006.

Apart from global warming, the anomalously high temperatures in Europe in autumn 2006 were mainly caused by the linear effects of a persistent southerly wind direction advecting warm air to the north and persistence from the very hot July along the shores of the North Sea.  A simple linear model incorporating these factors reproduces most of the temperature anomalies west of the Alps, but not in Germany.

Current climate models already underestimate the observed mean warming in Europe relative to global warming before 2006.  They also do not show an extra increase of the warm tail of the distribution as the climate warms.  Either autumn 2006 was a very rare event, or these climate models do not give the correct change in temperature distribution as the temperature rises.

\begin{acknowledgements}
I would like to thank my colleagues at KNMI for helpful comments.  This work was partly supported by the European Commission's 6th Framework Programme project ENSEMBLES (contract number GOCE-CT-2003-505539).  The ESSENCE project, lead by Wilco Hazeleger (KNMI) and Henk Dijkstra (UU/IMAU), was carried out with support of DEISA, HLRS, SARA and NCF (through NCF projects NRG-2006.06, CAVE-06-023 and SG-06-267). We thank the DEISA Consortium (co-funded by the EU, FP6 projects 508830 / 031513) for support within the DEISA Extreme Computing Initiative (www.deisa.org). The author thank Andreas Sterl (KNMI), Camiel Severijns (KNMI), and HLRS and SARA staff for technical support.  We acknowledge the other modelling groups for making their simulations available for analysis, the Program for Climate Model Diagnosis and Intercomparison (PCMDI) for collecting and archiving the CMIP3 model output, and the WCRP's Working Group on Coupled Modelling (WGCM) for organising the model data analysis activity.  The WCRP CMIP3 multi-model dataset is supported by the Office of Science, U.S. Department of Energy.  All data used is available on the KNMI Climate Explorer \citep[climexp.knmi.nl;][]{vanOldenborghdecadal}.
\end{acknowledgements}


\end{document}